\documentclass[12pt]{JHEP3}
\usepackage{mathrsfs}
\usepackage{amsmath,amssymb}
\usepackage{epsfig}
\usepackage{relsize}
\usepackage{graphicx}
\usepackage{float}

\def\({\left(}
\def\){\right)}

\def\sl(2){\alg{sl}(2)}

\def\be{\begin{equation}}
\def\ee{\end{equation}}

\newcommand{\bea}{\begin{eqnarray}}
\newcommand{\eea}{\end{eqnarray}}

\def\pa {\partial}

\def\la{\label}

\def\ov{\over}

\def\tp{{\widetilde p}}

\newcommand{\alg}[1]{\mathfrak{#1}}
\newcommand{\su}{\alg{su}}

\newcommand{\AdS}{{\rm  AdS}_5\times {\rm S}^5}

\newcommand{\bem}{\left (\begin{matrix}}
\newcommand{\eem}{\end{matrix} \right )}

\def\lam{\lambda}

\author{
Sergey Frolov\footnote{Email: frolovs@maths.tcd.ie} {}\footnote{Correspondent fellow at Steklov
Mathematical Institute, Moscow.}
\\  {\it Hamilton Mathematics Institute and School of Mathematics, \\
~~Trinity College, Dublin 2, Ireland} }

\abstract{  TBA equations for two-particle states from the $\sl(2)$ sector proposed by Arutyunov, Suzuki and the author are solved numerically for the Konishi operator descendent up to 't Hooft's coupling $\lambda\approx 2046$. 
The data obtained is used to analyze the properties of Y-functions and address the issue of the existence of the critical values of the coupling. 
In addition we  find a new integral representation for the BES dressing phase which substantially reduces the computational time.
}

\title{Konishi operator at intermediate coupling
}

\preprint{
          \smaller{\smaller{\smaller{TCDMATH 10-05}}}\\[-.5ex]
          \smaller{\smaller{\smaller{HMI-10-03}}}
          }

\begin{document}

\renewcommand{\thefootnote}{\arabic{footnote}}
\setcounter{footnote}{0}


\section{Introduction}

The Thermodynamic Bethe Ansatz (TBA) is an efficient tool to determine the finite-size spectrum of two-dimensional relativistic integrable models \cite{Zamolodchikov90}. The relevance of the TBA for the $\AdS$ spectral problem was advocated in  \cite{AJK} where  L\"uscher's approach \cite{Luscher85} was used to relate exponential corrections to string energy to wrapping effects in dual field theory.  Its application to a nonrelativistic model requires understanding thermodynamic properties of a so-called mirror theory
\cite{AF07} which is obtained from the original model by means of a double-Wick rotation.  The mirror model of the light-cone  $\AdS$ superstring was studied in detail in \cite{AF07} where, in particular, the asymptotic spectrum was identified, and the mirror form of the Bethe-Yang equations of \cite{BS} was determined. 

The major step towards realizing the TBA approach is to formulate a so-called string hypothesis  \cite{Takahashi72} which 
classifies the states contributing in the thermodynamic limit.  TBA equations and the associated Y-system are then readily derived from it, see e.g. \cite{Korepin}.  This step was made 
last year in  \cite{AF09a} where the mirror Bethe-Yang equations \cite{AF07} were used to formulate a string hypothesis for the $\AdS$ mirror model.  This opened a way to derive the corresponding canonical \cite{AF09b,BFT,GKKV09} and simplified TBA equations \cite{AF09d}.

The TBA equations combined with a certain analytic continuation procedure proposed for relativistic models in  \cite{DT96,DT97} can be also used to find energies of excited states.\footnote{For other approaches and applications, see e.g. \cite{BLZe}-\cite{GKV08}.} The procedure was called the contour deformation trick in \cite{AFS09} because 
it basically reduces to deforming the integration contours in ground-state TBA equations while keeping their form  untouched. As a result,  excited states TBA equations differ only by  integration contours.
For practical applications however one should take the integration contours back to their ground-state positions, and this results in the appearance of state-dependent driving terms in the TBA equations. 

The contour deformation trick was used in  \cite{AFS09} to analyze two-particle states from the $\sl(2)$ sector. It was shown there that two-particle states are divided into infinitely-many classes, each class having its own set of 
driving terms in the TBA equations. 
For the Konishi-like states the TBA equations of \cite{AFS09}  are believed to be equivalent to those of \cite{GKKV09, GKV09b},\footnote{The set of integral equations proposed in  \cite{GKKV09, GKV09b} was named an integral form of the Y-system. We find this terminology somewhat misleading because these integral equations were not derived from the Y-system conjectured in  \cite{GKV09} but were proposed by following the pure TBA approach, that is by using the canonical TBA equations  \cite{AF09b,BFT,GKKV09}, the contour deformation trick  \cite{DT96}, and the large $J$ asymptotic solution of the Y-system  \cite{GKV09}.  A derivation of the TBA equations from the Y-system requires understanding the complicated analytic structure  of Y-functions, see \cite{AF09b,FS,AFS09, CFT10} for some results in this direction.}  and our numerics strongly supports this.

The TBA equations obtained in  \cite{AFS09} were used to  derive the five-loop anomalous dimension of the Konishi operator, and to show numerically  \cite{AFS10}  that the corresponding result perfectly agrees with the one recently obtained via the generalized L\"uscher formulae \cite{BJ09}. 
The analysis in \cite{AFS09,AFS10} was then extended 
to prove the agreement analytically first  for the Konishi operator \cite{BH10a}, and then for an arbitrary twist-two operator  \cite{BH10b} reproducing the results in \cite{LRV09}.\footnote{The five-loop computations in \cite{BJ09,LRV09} were based on \cite{BJ08} where the four-loop anomalous dimension of the Konishi operator was computed and shown to agree with the direct field theory result \cite{Sieg,Vel} . This constituted a strong test of the AdS/CFT correspondence \cite{M}. For other applications of L\"uscher's approach, see 
e.g. \cite{JL07}-\cite{Vel10}.} For other applications of the TBA approach to the AdS/CFT spectral problem, see  \cite{Correa}-\cite{Gromov09b}.

In a parallel development the TBA 
equations proposed in  \cite{GKKV09} were used in \cite{GKV09b} to compute anomalous dimension of the Konishi operator up to a relatively large value of the  't Hooft coupling constant $\lambda\approx 664$. Analyzing the results obtained, 
the authors of \cite{GKV09b}  found the following fitting function for the Konishi state energy or,  equivalently, for the conformal dimension of the dual Konishi operator
\be\label{EGKV}
{\overline E}_K^{GKV}(\lam)=\sqrt[4]{{\lam}}\left(2.0004+\frac{1.988}{
   \sqrt{{\lam}}}-\frac{2.60}{\lam}+\frac{6.2}{\lam^{3/2}}\right)\,.
\ee
This formula allows one to make four remarkable predictions. First, it predicts that the coefficient $c_{-1}$ of the leading term in the large $\lam$ expansion is equal to 2 which agrees with the spectrum of string theory in flat space \cite{GKV02} and asymptotic Bethe ansatz considerations \cite{AFS}. Second, it shows the vanishing of the constant term $c_0$.
Third,
it makes a new prediction which disagrees with the semiclassical consideration in \cite{RT}  that the first nonvanishing subleading coefficient is also equal to 2. And fourth, it predicts that up to an overall factor of $\sqrt[4]{{\lam}}$ the Konishi state energy is a series in $1/\sqrt{{\lam}}$.
The last prediction in fact  agrees with the argumentation  in \cite{RT}  but it disagrees with the considerations in \cite{AF05} where 
the free fermion model describing the $\su(1|1)$ sector in the semi-classical approximation was analyzed, and it was shown that the strong coupling expansion for short states is in powers of  $1/\sqrt[4]{{\lam}}$. It is worth noting that this simple model indeed predicts that the constant term in the strong coupling expansion vanishes.  The formulae derived in the framework of the free fermion model would definitely get quantum corrections, and if the prediction of \cite{GKV09b} is correct it would imply that these corrections 
drastically change the structure of the strong coupling expansion.

In this paper we reconsider the computation of the Konishi state energy. We  solve numerically
the excited states TBA equations \cite{AFS09}  for the Konishi operator up to 't Hooft's coupling $\lambda\approx 2046$, and use the data obtained to analyze the behavior of Y-functions. 
We use the analysis to address the issue of the existence of the critical values of $\lambda$ raised in \cite{AFS09}. The consideration in \cite{AFS09} was based on an assumption that the analytic properties of the exact Y-functions would follow those of the large $J$ asymptotic solution, and if this assumption is not realized then the TBA equations of \cite{GKV09b, AFS09} formulated for Konishi-like states at weak coupling may be valid  for all values of $\lambda$. Extrapolating our results 
we find that the first critical value for the Konishi operator is 
most probably greater than 5300, $\lambda_{cr}^{(1)} >5300$,
 which is significantly higher than the estimate based on the
large $J$ asymptotic solution that gives $\lambda_{cr}^{(1)} \approx774$. 

We also find that
the contribution of $Y_Q$-functions to the Konishi state energy grows almost linearly with the string tension $g=\sqrt\lambda/2\pi$. If this behavior continues to hold for all values of $\lambda$ then this would imply that at large $\lambda$ the exact Bethe root asymptotes to a constant less than 2. This would be drastically different from the asymptotic behavior of the corresponding solution of the Bethe-Yang equations and would, in particular, imply the absence of  the critical values for the Konishi operator. 
This would be a very puzzling scenario because 
 the full spectrum of string theory in flat space can be reproduced  already from the Bethe-Yang equations  \cite{AFS}, and if $w$ does not asymptote to 2 it would be necessary to explain how the spectrum follows from the TBA equations.
In particular, the spectrum degeneracy would be more difficult to explain  because Bethe roots of different states would not behave uniformly at strong coupling.

Examining  the data obtained, 
we found convincing evidence  in favor of the prediction of \cite{GKV09b} that
 the first nonvanishing subleading coefficient $c_1$ of the $1/\sqrt[4]\lam$ term is equal to $2$, and the coefficient $c_2$ of the terms $1/\sqrt\lam$ vanishes. 
 If one cuts the asymptotic series at order $1/\lam^{5/4}$ and fits our data then the coefficient $c_1$ appears to be $2.02\pm 0.02$ depending on the fitting interval used. It is clearly reasonable to assume that $c_1$ is equal to $2$. 
Setting $c_1=0$, one then finds that $c_2= -0.02\pm 0.01$ which is obviously very small. 
 
Then, assuming from the very beginning that $c_{-1}=c_1=2$ and  $c_0=c_2=0$
 and fitting the data with $\lam>77$, one gets the following fitting function for the Konishi state energy
 \bea
\la{FFit14i}
c_{-1}=c_1=2\,,\ c_2=0\ \ \Longrightarrow\ \   {\overline E}_K(\lam)= 2
   \sqrt[4]{\lam}+\frac{2}{\sqrt[4]{\lam}}-\frac{3.26}{\lam^{3/4}}+\frac{2.53}{\lam}+
 \frac{4.03}{\lam^{5/4}}\,,~~~~~
 \eea
which differs from \eqref{EGKV} by the presence of the $1/\lam$ term. 
Thus, the coefficient $c_4$ of the term $1/\lam$ does not vanish, and 
we cannot confirm the prediction of 
\cite{GKV09b} that up to an overall
factor of $\sqrt[4]\lam$  the large $\lam$ expansion of the Konishi state energy is a series in $1/\sqrt\lam$. The vanishing of $c_2$ could be related to the high degree of supersymmetry of the model as was pointed out in \cite{RT}.

If on the other hand one follows \cite{GKV09b} and assumes from the very beginning that $c_2=c_4=0$
 then fitting the data with $\lam>77$, one gets the following fitting function for the Konishi state energy
\be\la{Emy}
c_2=c_4=0\ \ \Longrightarrow\ \  {\overline E}_K(\lam)= \sqrt[4]{\lambda
   }\left(2.00005+\frac{1.992}{\sqrt{\lambda }} -\frac{2.73}{\lambda}+\frac{7.45}{\lambda ^{3/2}}\right)\,,~~
\ee
which obviously is in a very good agreement with \eqref{EGKV}.
The coefficients in \eqref{Emy} mildly depend on the fitting interval used and, for example, the last coefficient can take values from 6 to 8. 

Let us also mention that our numerical results agree with those of \cite{GKV09b}    
with the $0.0015$ precision  for most values of $\lam$, and this implies the equivalence of  the TBA equations of \cite{GKKV09} and \cite{AFS09} for Konishi-like states at weak coupling.\footnote{Since the considerations in  \cite{AFS09} and in v.3 of \cite{GKKV09} have the same starting point -- the mirror theory string hypothesis \cite{AF09a} --  the equivalence of the TBA equations in fact follows from the equality of the mirror theory dressing phases used in  \cite{GKV09b} and  \cite{AFS09}
which was recently proven in \cite{CFT10}.
}

We used in our computation  a new integral representation for the BES dressing phase \cite{BES} which significantly reduces the computational time. 

The paper is organized as follows. In the next section we present
the results of the numerical solution of the TBA equations for the Konishi state energy. In section 3 we discuss the properties of Y-functions and estimate the first critical value of $\lam$.  In section 4 we use our data to find the coefficients of the large $\lam$ asymptotic expansion.
In  the appendix  we describe the numerical algorithm used for the computation, and present new formulae for  various dressing phases and kernels.

\section{Konishi state energy}

We solve numerically the following equations from \cite{AFS09}:  the simplified TBA equations (4.2-4.5)
for $Y_{M|w}$, $Y_{M|vw}$, and $Y_{\pm}$--functions,  and the hybrid equations (4.13) for $Y_{Q}$-functions, and determine the values of the Bethe root $u_2=-u_1\equiv w$ or, equivalently, the momentum $p=p(w)$ carried by a string particle from the exact Bethe equation (8.61).  The energy of the Konishi state is then given by the following formula
\be\la{EKon}
E_K(\lam)= 2+2 \sqrt{1+4g^2\sin^2{p\ov 2}} -\sum_{Q=1}^\infty\int\, {d\tp\ov 2\pi}\, \log(1+Y_Q)\,,
\ee
where $\tp$ is the momentum of a $Q$-particle of the mirror theory, and $g$ is the string tension related to 't Hooft's coupling $\lam$ as $\lam=4\pi^2g^2$.
The energy is given by the difference of the two terms -- the first one is the contribution coming from the dispersion relation, and the second one is the $Y$-functions contribution.  We denote the contributions as $E_{\rm dis}$ and $E_{\rm Y}$, respectively, so that $E_K= E_{\rm dis} - E_{\rm Y}$. It is clear that $E_{\rm dis}$ and $-E_{\rm Y}$ play the roles of the kinetic and potential energy of the Konishi state particles. We will also use the notation $E_{\rm Y_Q}$ for the contribution of a $Y_Q$-function to the energy.

It is known that at large values of  't Hooft's coupling the energy of the Konishi state can be expanded in an asymptotic series in powers of $1/ \sqrt[4]{{\lam}} $
\be
\la{asexp}
E_K(\lam)=2 \sqrt[4]{{\lam}}+c_0+\frac{c_1}{
   \sqrt[4]{{\lam}}}+\frac{c_2}{\sqrt{{\lam}}}+\frac{c_3}{\lam^{3/4}}+\frac{c_4}{\lam}+\frac{c_5}{\lam^{5/4}}+ \cdots\,,
   \ee
where the leading $2 \sqrt[4]{{\lam}}$ behavior follows from the string spectrum in flat space  \cite{GKV02}, and can be reproduced from the asymptotic Bethe ansatz \cite{AFS}. The coefficient $c_0$ is believed to be equal to 0 because that is what one gets from both the free fermion model describing the $\su(1|1)$ sector and  the asymptotic Bethe ansatz \cite{AF05} but to our knowledge an honest string theory derivation of  $c_0$ is absent.  The large $\lambda$ perturbative expansion of the light-cone string sigma model (see \cite{AFrev} for a review) allows one to have any of the coefficients $c_i$ nonvanishing. 
It was however argued in \cite{RT} that the coefficient  $c_2$ should vanish due to the high degree of supersymmetry of the model. 
Moreover, it follows from the fitting function \eqref{EGKV}  for the Konishi state energy obtained by solving the canonical TBA equations \cite{GKV09b} that 
in fact both the coefficients  $c_2$ and $c_4$ vanish, and then it is tempting to speculate that
all coefficients with even indices vanish: $c_{2k}=0$. 

One of our aims is to demonstrate that the excited state TBA equations of \cite{AFS09} indeed predict that the leading coefficient is 2,  and then to understand if one can fix the coefficients $c_k$, and in particular if the numerical data indeed predicts that $c_0=c_2=c_4=0$. 

\begin{figure}[t]
\begin{center}
\includegraphics*[width=0.6\textwidth]{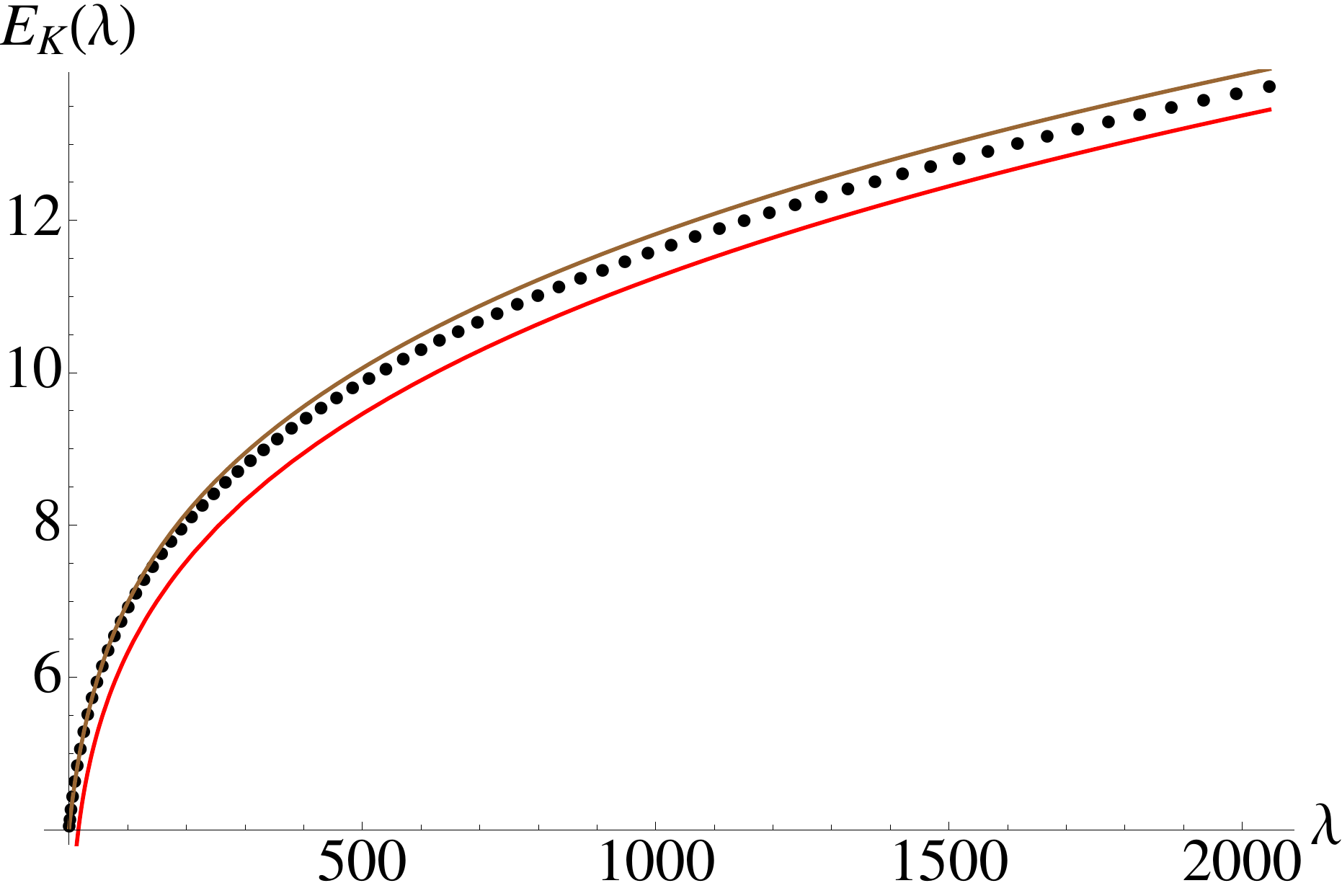}
\end{center}
\caption{\smaller 
Black dots represent the numerical solution of the TBA equations for the Konishi state energy $E_K(\lambda)$. The brown (upper) curve represents the solution of the Bethe-Yang equation, and the red (lower) curve  is the graph of $2 \sqrt[4]{{\lam}}$ which is  the  large $\lam$ asymptote of ${E}_K(\lam)$. 
The range of the coupling constant is from $g=0.1, \lam=0.39$ to $g=7.2, \lam=2046.56$. 
}
\end{figure}

The results of our computation of the Konishi state energy or, equivalently, the conformal dimension of the Konishi operator as a function of $g$ are collected 
in the table \eqref{Edata} from the Appendix. In Figure 1 we plot the data together with   the graph of the function $E_{\rm dis}(w_{\rm asym})$ where $w_{\rm asym}=w_{\rm asym}(\lam)$ is the corresponding solution of the Bethe-Yang equation, and the graph of $2 \sqrt[4]{{\lam}}$ which is the large $\lam$ asymptote of both the exact and asymptotic  energies. 
\noindent One sees that the exact energy graph approaches $2 \sqrt[4]{{\lam}}$ faster than $E_{\rm dis}(w_{\rm asym})$. It is also useful to plot the graph of  
$E_{\rm dis}(w)$ as a function of $\lam$ to compare its contribution with the exact energy. The corresponding plots are shown in Figure 2 where we use the string tension $g=\sqrt\lam /2\pi$ as an independent variable. We see that $E_{\rm dis}(w)$ grows almost linearly with $g$ for $g>4$ while the total energy grows as $\sqrt g$. 
\begin{figure}[t]
\begin{center}
\includegraphics*[width=0.6\textwidth]{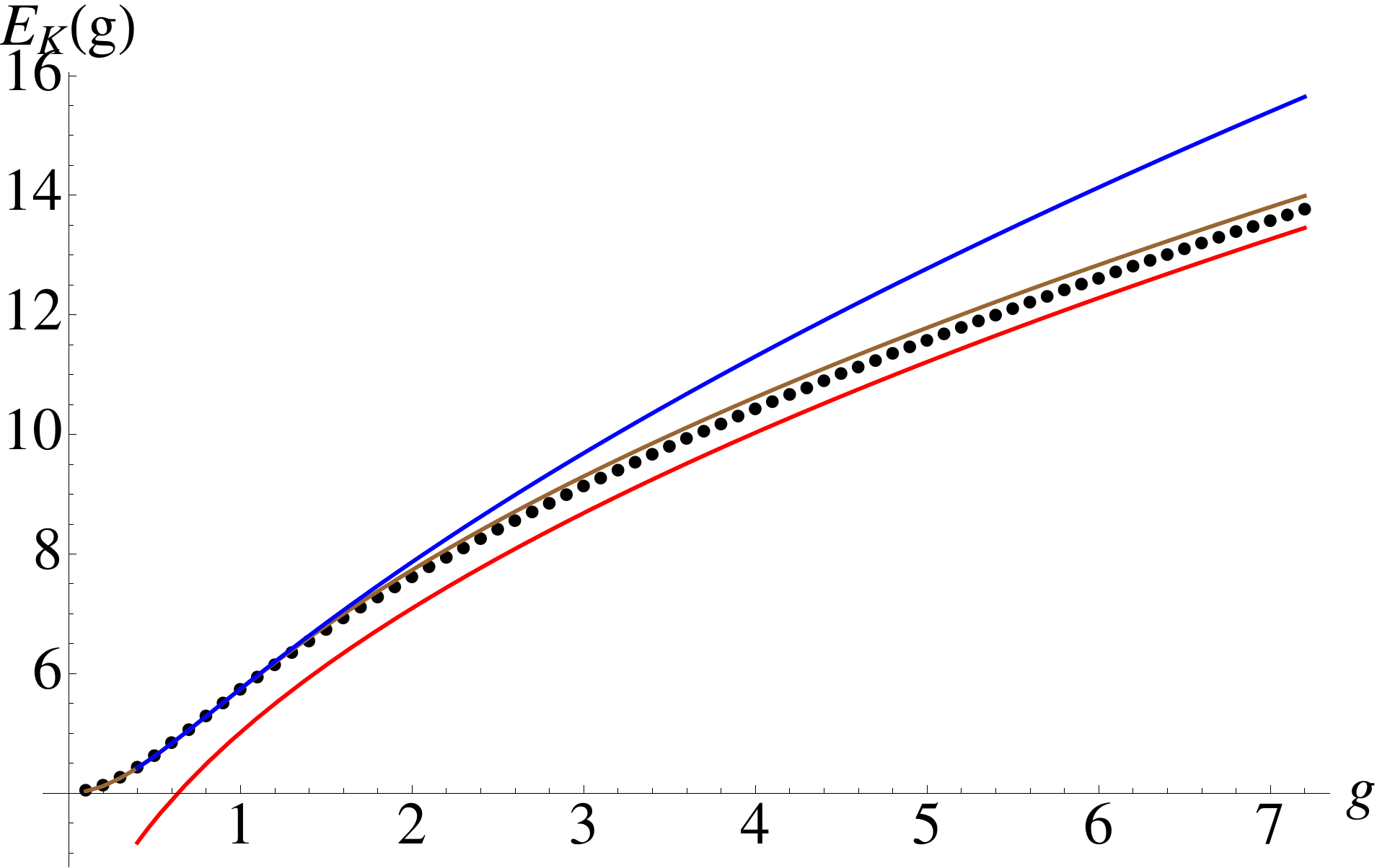}
\end{center}
\caption{\smaller Here black dots, the brown and red  curves are the same as in Figure 1, and the blue (upper) curve is the graph of $E_{\rm dis}(w)$ where $w=w(g)$ is the solution of the exact Bethe equations. }
\end{figure}
This implies immediately that the Y-functions contribution also grows linearly, see Figure 3, and, moreover, the linear parts of $E_{\rm dis}(w)$ and $E_{\rm Y}$ cancel each other leading to the $\sqrt g$ behavior of $E_K$.   
\begin{figure}[t]
\begin{center}
\includegraphics*[width=0.6\textwidth]{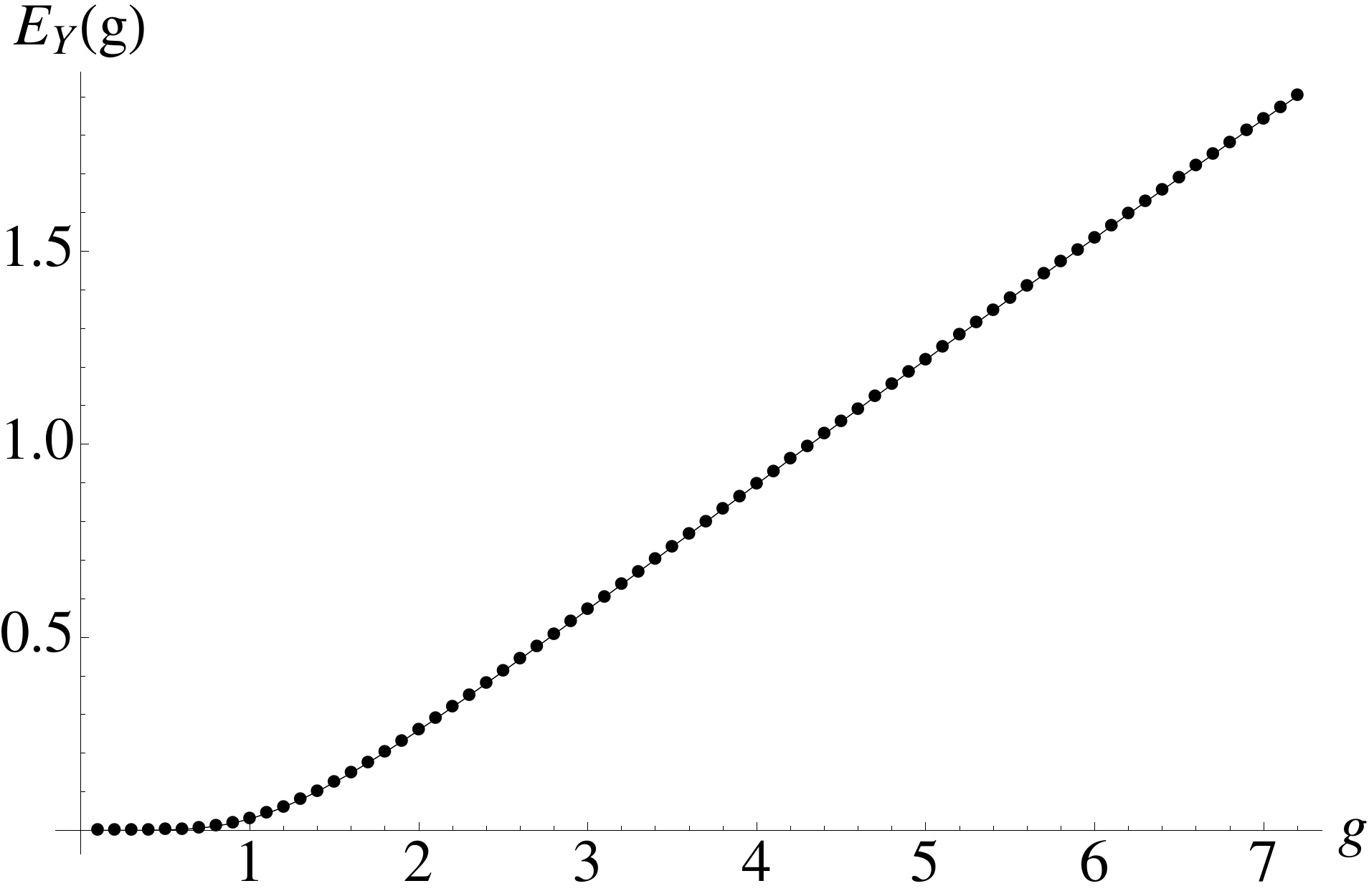}
\end{center}
\caption{\smaller This is the graph of the contribution of $E_{\rm Y}$ to the energy. It obviously shows a linear growth starting already with $g\sim 2$. }
\end{figure}
On the other hand,  $E_{\rm dis}(w_{\rm asym})$ grows only as $\sqrt g$ for these values of $g$. It is interesting that the linear dependence of $E_{\rm Y}$ becomes clearly visible already at very small values of $g$. To understand the reason for such a different behavior of $E_{\rm dis}(w)$ and $E_{\rm dis}(w_{\rm asym})$ we plot 
in Figure 4 the solutions of the exact Bethe equation  and the Bethe-Yang equation for $w$ and $w_{\rm asym}$, respectively. The numerical data of the computation of the Bethe root $w$ are in the table \eqref{wdata} in Appendix.

\begin{figure}[t]
\begin{center}
\includegraphics*[width=0.6\textwidth]{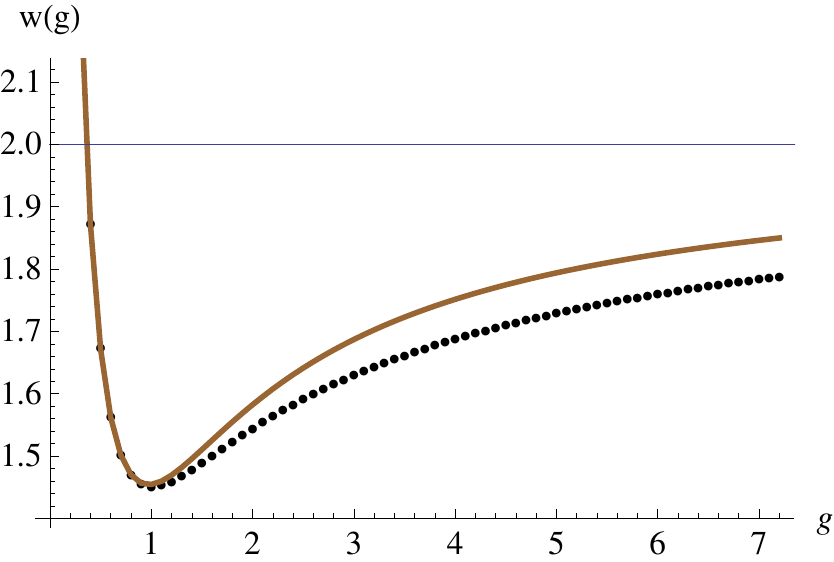}
\end{center}
\caption{\smaller The black dots represent our numerical solution of the exact Bethe equation, and the brown curve is the graph of  the corresponding solution of the Bethe-Yang equation.
The exact Bethe root $w(g)$ reaches its minimum at $g\approx 1$, and  $g\approx 1.6$ is the inflection point. 
}
\end{figure}
\noindent One sees that $w$ is still pretty far from 2 which is the large $g$ asymptotic value of $w_{\rm asym}$, and that the exact Bethe root is noticably smaller than  $w_{\rm asym}$. The corresponding exact momentum $p$ is not small at these values of $g$ and decreases much slower than $p_{\rm asym}$. This explains why the dispersion relation  contribution indeed grows as $g$.

It is tempting to conclude from Figures 2 and 3 that  $E_{\rm dis}(w)$ and $E_{\rm Y}$ would grow linearly for all values of $g$, and, therefore, $w$ in fact would asymptote to a constant $w_\infty$ less than 2. If this happens then 
 this would make the existence of 
critical values of the coupling constant discussed in \cite{AFS09} rather improbable,
and as a result the TBA equations for Konishi-like states \cite{GKV09b,AFS09} might  be valid for any value of $g$.  Also, this would mean that the strong coupling limit of multi-particle states with finite  number of particles and $J$ (short operators in dual field theory) is very different from the near flat space limit discussed in \cite{AFS,MS} where the rapidities asymptote to 2, and one can study states with $J\sim \sqrt g$. The puzzle then is that the full spectrum of string theory in flat space can be reproduced  already in the near flat space limit \cite{AFS}, and if $w$ does not asymptote to 2 it would be necessary to explain how the flat space string spectrum follows from the TBA equations.\footnote{Strictly speaking, even if $w$ asymptotes to 2 but with a rate different from the one of $w_{\rm asym}$ it would be a challenge to derive the 
 flat space string spectrum from TBA.} 
 
This is an intriguing scenario, and 
it would be very interesting to understand analytically if it is the one. Our numerics however seems to indicate that the linear growth of  $E_{\rm Y}$ might be a feature of the intermediate coupling regime we are studying, and it will slow down for larger values of $g$. 
In Figure 5 we plot the graphs of the derivative of $E_{\rm dis}(w)$  and $E_{\rm Y}$  with respect to $g$ (obtained by using the Interpolation function in Mathematica).  We see that the rate of change of $E_{\rm Y}$ reaches its maximum at $g\sim 3$, remains almost constant till $g\sim 4$  and then begins to decrease very slowly. We are not sure if this effect is genuine. The precision of our computation falls down for $g>4$, and the decrease in the rate of change of $E_{\rm Y}$ may be just a numerical artifact. Since, as will be discussed in the next section, the contribution of an individual $Y_Q$--function to $E_{\rm Y}$ slightly decreases at large $g$ it might be also necessary to include the contribution of more $Y_Q$--functions than we did. Then,
it is certainly possible that the rate of change of $E_{\rm Y}$ would stabilize at even larger values of $g$, and the scenario discussed above would be realized. 
\begin{figure}[t]
\begin{center}
\includegraphics*[width=0.45\textwidth]{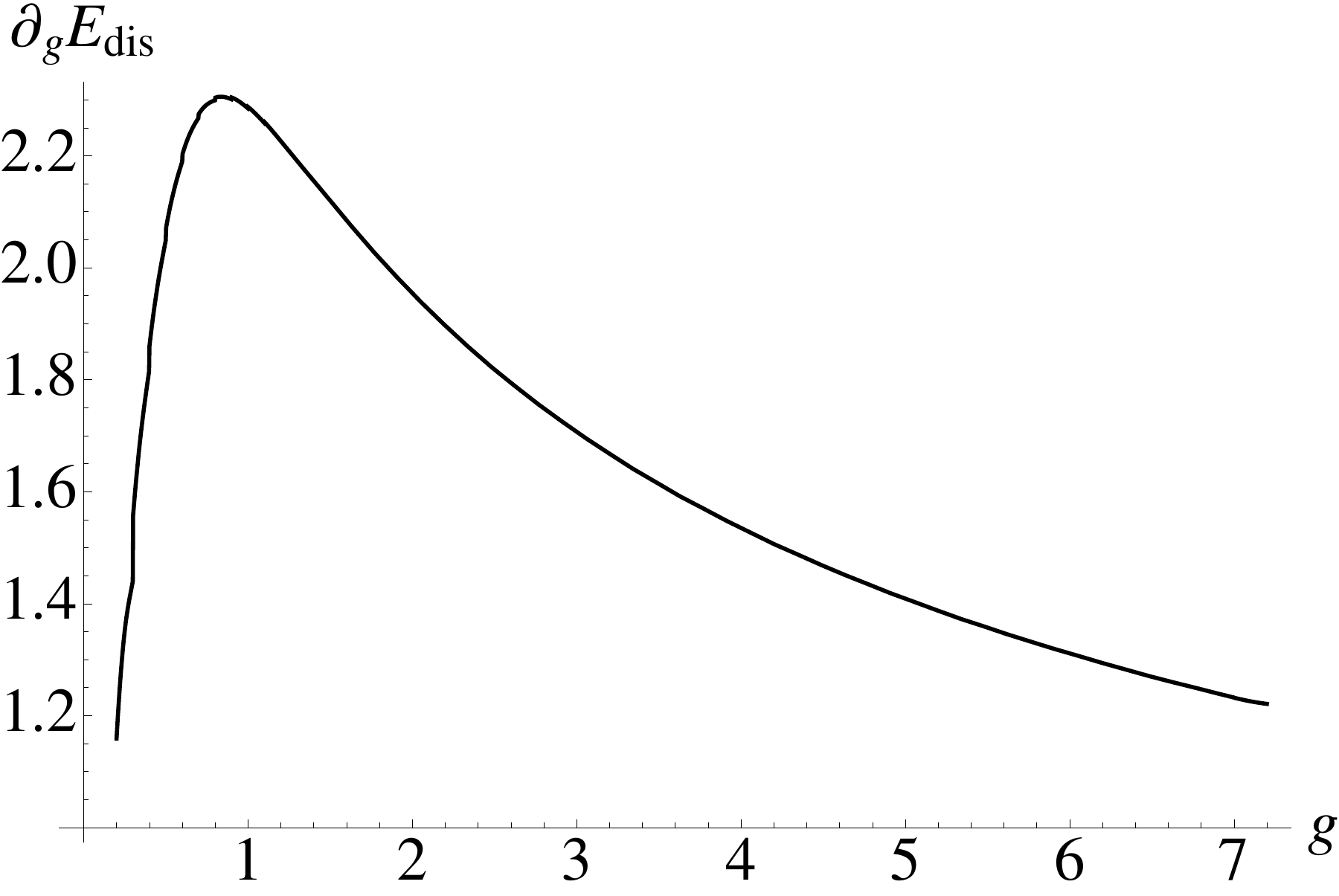}\quad \includegraphics*[width=0.45\textwidth]{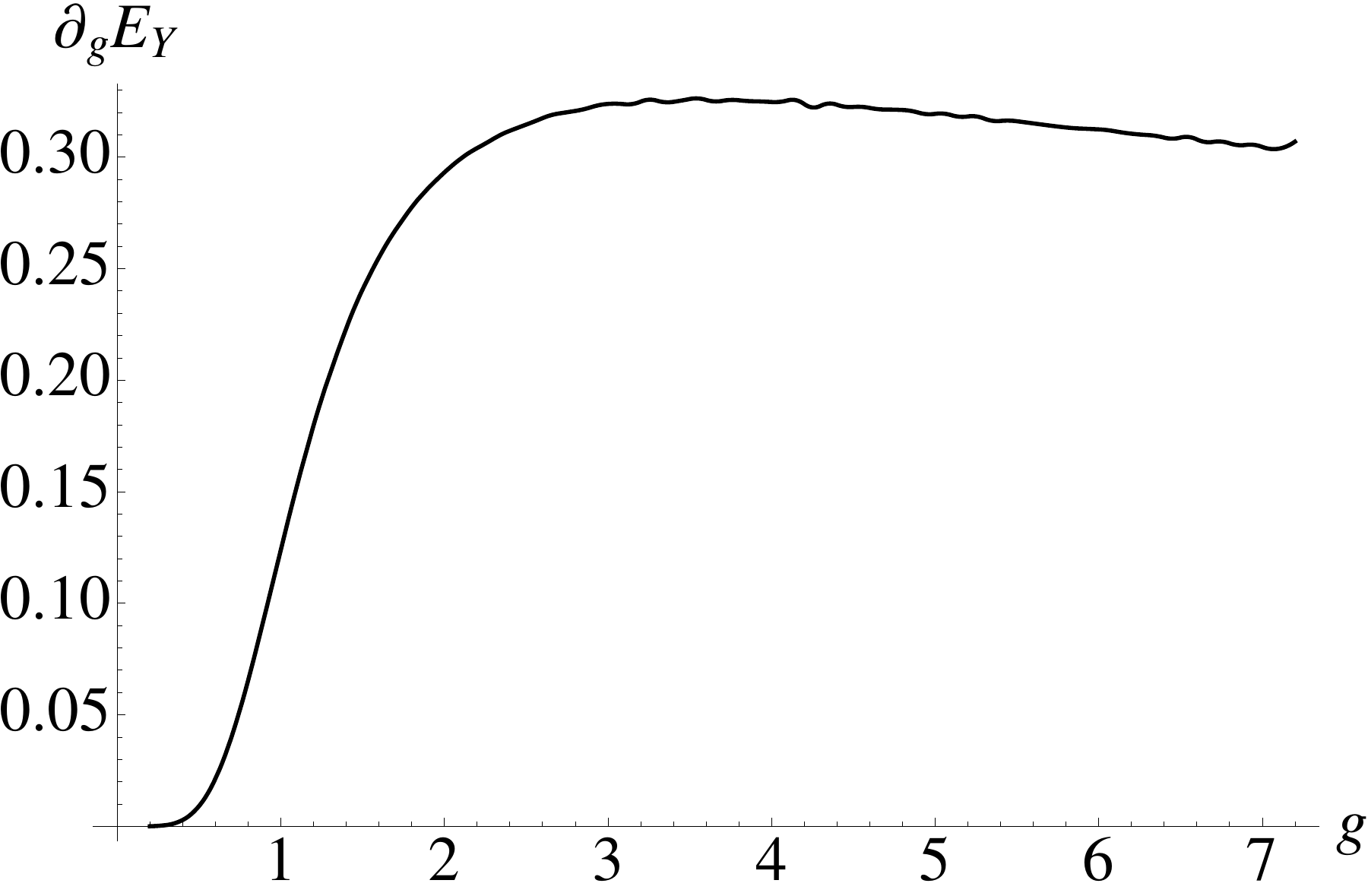}
\end{center}
\caption{\smaller The graphs of the derivative of  $E_{\rm dis}$ and $E_{\rm Y}$ with respect to $g$.
}
\end{figure}
The assumption that the Y-functions contribution  would be linear in $g$ for very large $g$ has a profound consequence on the strong coupling dependence of the $Y_1$--function. 
Since $Y_Q$-functions are very small for $|u|>2$ the integration region in \eqref{EKon} is effectively of order $\sqrt g$ for large $g$, and, therefore, the Y-functions contribution would grow as $g$ only if  $\log Y_1$ would be of order $\sqrt g$.  However, in the next section we will see that it is not the case and for $g\sim 7$, $Y_1$ increases only as $g^{3/2}$. This behavior is different from both the asymptotic $Y_1$--function $g$-dependence
and the exponential growth required by the scenario discussed above. 
This also shows clearly that the values of $g$ we have reached are not large enough, and we are still analyzing the intermediate coupling regime.
What happens at larger values of $g$ remains to be understood. 

\section{Y-functions}

In this section we discuss various properties of Y-functions. We begin with $Y_Q$-functions because the energy of the Konishi state depends explicitly on them.

In Figure 6 we show plots\footnote{All Y-functions are even so we plot them only for  $u\ge 0$.} of several exact $Y_1$-functions computed at various values of $g$. 
\begin{figure}[t]
\begin{center}
\includegraphics*[width=0.45\textwidth]{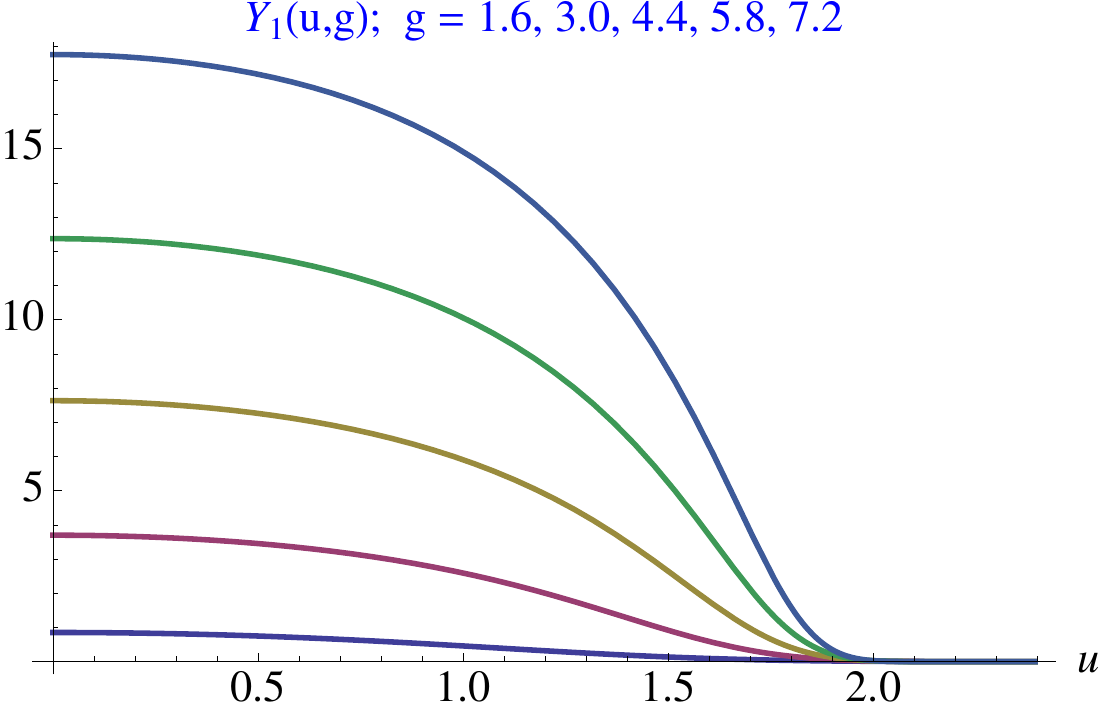}\quad \includegraphics*[width=0.45\textwidth]{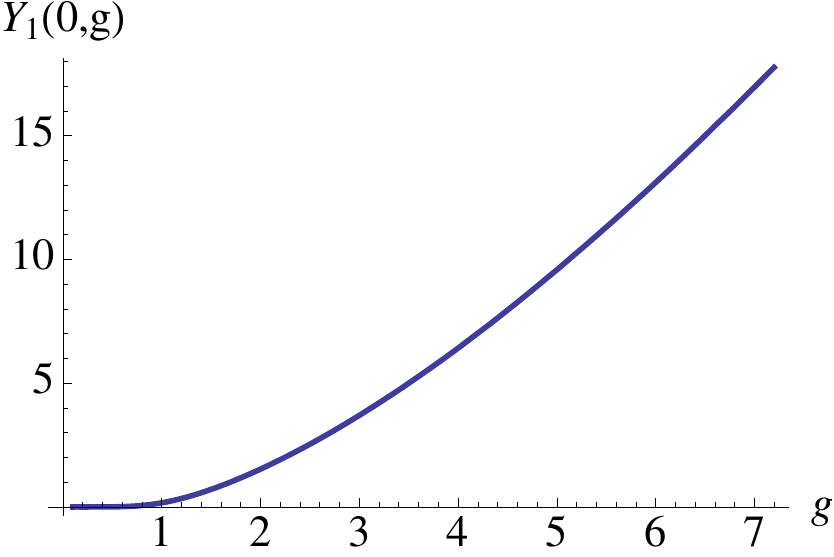}
\end{center}
\caption{\smaller  On the left figure graphs of five $Y_1$-functions with different values of $g$ are shown, and on the right figure the graph of $Y_1(0,g)$ as a function of $g$ is plotted.
}
\end{figure}
One sees that $Y_1$ is getting larger in the interval $[-2,2]$ with $g$ increasing. In fact it increases very fast, and becomes much larger (one order of magnitude) than the asymptotic $Y_1^o$-function computed at the same values of $g$ and $w$.  
In particular, $Y_1(0,g)$ keeps increasing while the asymptotic $Y_1^o$-function computed with the exact Bethe root $w$ decreases  at $u=0$ for $g>4$. To find the $g$-dependence we  plot $\sqrt g\, Y_1(0,g)$ and $\sqrt g\, Y_1(1,g)$ in Figure 7. We see that they are almost linear functions, and, therefore, $Y_1\sim g^{3/2}$.
\begin{figure}[t]
\begin{center}
\includegraphics*[width=0.45\textwidth]{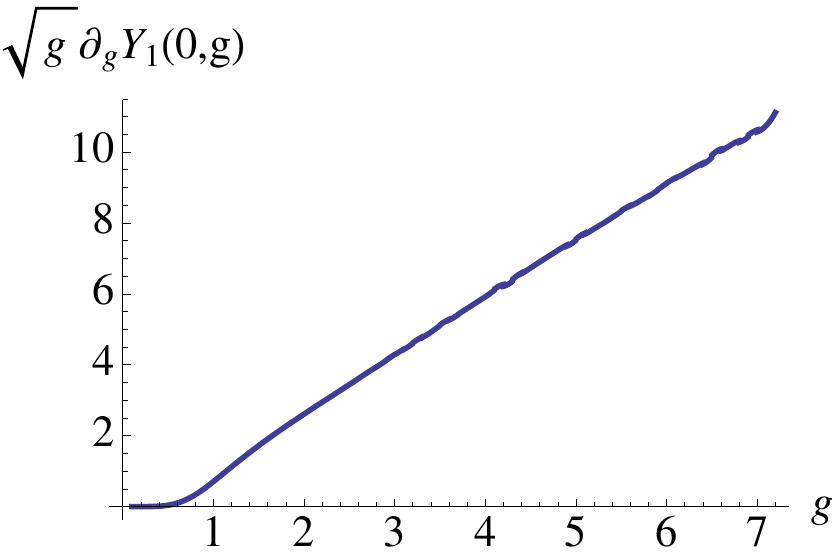}\quad \includegraphics*[width=0.45\textwidth]{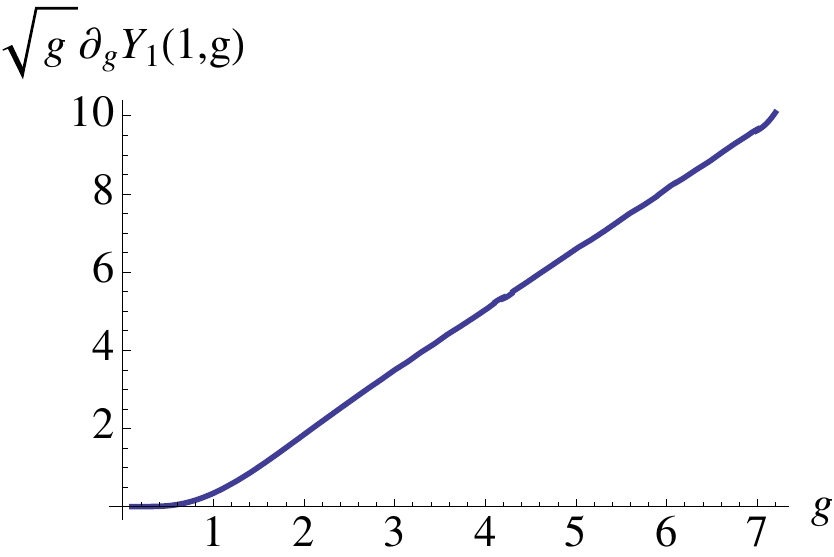}
\end{center}
\caption{\smaller  The graphs of $\sqrt g\pa_g Y_1(0,g)$ and $\sqrt g \pa_g Y_1(1,g)$ definitely show that $Y_1$ grows as $g^{3/2}$. 
}
\end{figure}
\noindent It is hardly possible that  $Y_1$ would show the $g^{3/2}$-dependence for large $g$ because this would lead to the contribution of the order $\sqrt g\log g$ to the Konishi state energy which cannot be canceled by any reasonable contribution  from $E_{\rm dis}$.  If the exact Bethe root $w$ approaches 2 for large $g$ then $Y_1(u,g)$ should  asymptote to a finite function. The existence of the critical values of $g$ seems to 
require in addition that $Y_1(u,g)$ would go to 0 for any $u<2$ (but it could stay finite for $u\sim 2-\nu/g$).  If $w$ approaches $w_\infty<2$ then, as was mentioned in the previous section, $\log Y_1$ must grow as $\sqrt g$. 
 
In Figure 8  the plot of the contribution $E_{\rm Y_1}$ of $Y_1$-function  to the Konishi state energy and the plot of  its derivative $\partial_gE_{\rm Y_1}$ are  shown. $E_{\rm Y_1}$ decreases too fast to be explained by insufficient numerical precision. The scenario with $w_\infty<2$ would be realized only if  $\partial_gE_{\rm Y_1}$ asymptotes to a positive constant.  
\begin{figure}[t]
\begin{center}
\includegraphics*[width=0.45\textwidth]{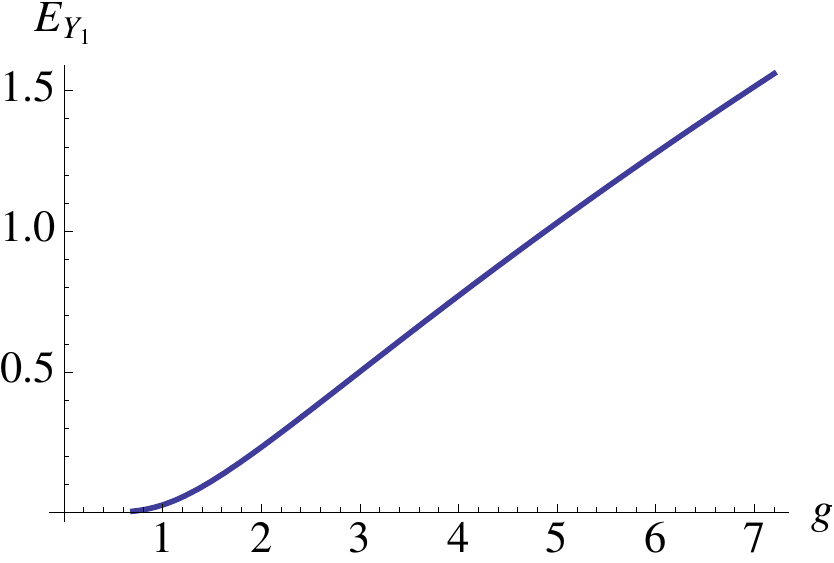}\quad \includegraphics*[width=0.45\textwidth]{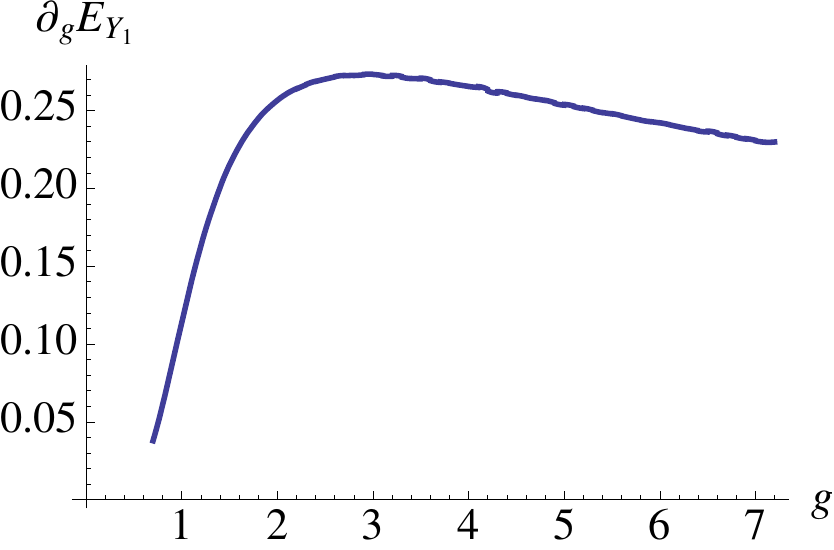}
\end{center}
\caption{\smaller  These are the plots of $E_{\rm Y_1}$ and $\partial_gE_{\rm Y_1}$. 
}
\end{figure}

Figure  9 shows  plots of  $Y_2$-functions.  One sees that they exhibit a rather intricate $g$ dependence. The maximum value of $Y_2$ increases with $g$ and shifts to the right towards $u=2$. 
\begin{figure}[H]
\begin{center}
\includegraphics*[width=0.45\textwidth]{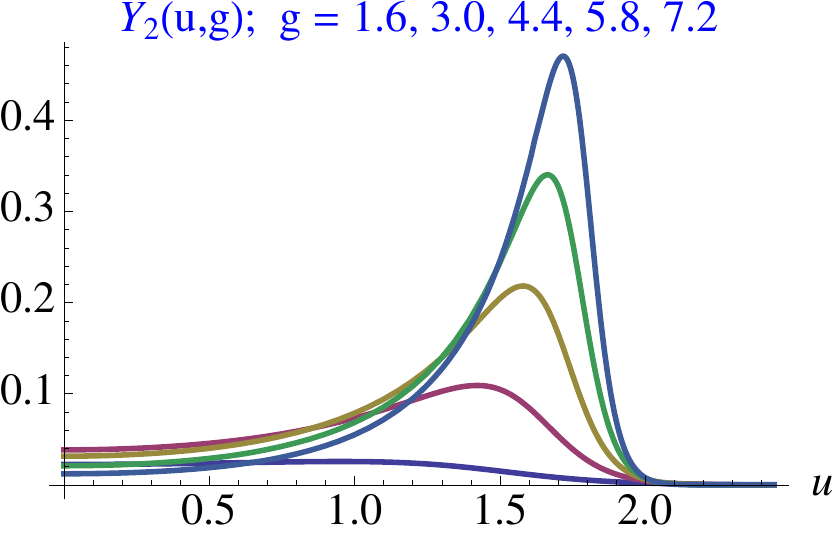}\quad \includegraphics*[width=0.45\textwidth]{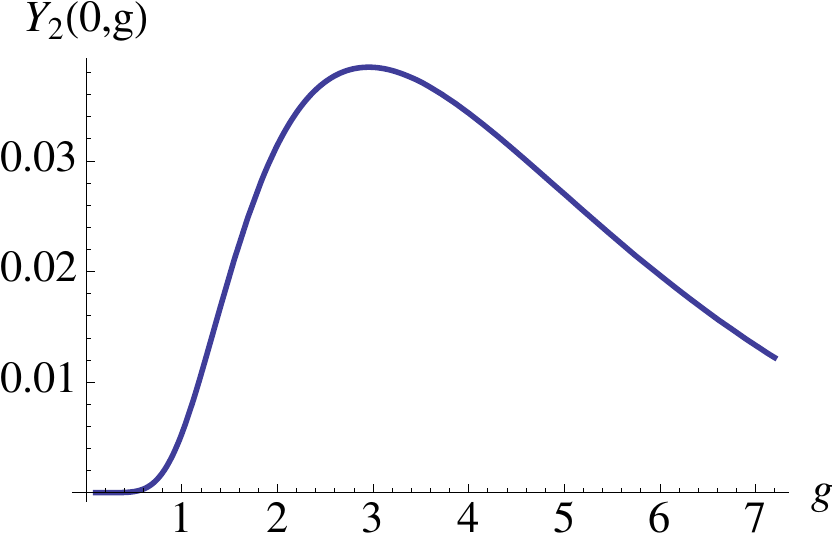}
\end{center}
\caption{\smaller  On the left figure graphs of five $Y_2$-functions with different values of $g$ are shown, and on the right figure the graph of $Y_2(0,g)$ as a function of $g$ is plotted. }
\end{figure}
We see from Figure 9 that $Y_2(0,g)$ is decreasing for  $g> 3$.
This behavior agrees with our expectations based on the analysis of asymptotic Y-functions.  Let us recall that $Y_2$-function is one of the Y-functions that can be used to find the first sub-critical value of $g$ because it vanishes at $u=0$ if $g=\bar{g}_{cr}^{(1)}$ \cite{AFS09}. Let us now assume that the critical value does exist. Then  since $Y_2$ is non-negative for all values of $g$ it should have an expansion of the form 
$$Y_2(0,g) \sim (g-\bar{g}_{cr}^{(1)})^2+\cdots\ .$$
Since most of our points are far from $\bar{g}_{cr}^{(1)}$, to estimate the first sub-critical value we first however use the linear fit. This would give the lowest estimate\footnote{We assume there will be no sharp changes in the behavior of $Y_2(0,g)$ for larger values of $g$. } of the value of $\bar{g}_{cr}^{(1)}$ because $Y_2(0,g)$ has a double zero  at  $g=\bar{g}_{cr}^{(1)}$. 
Fitting our data in the interval $[g_0,7.2]$ to the function $c_1 (g-\bar{g}_{cr}^{(1)})$, we get the results shown in \eqref{YQ2fit}.
{\smaller 
\bea\la{YQ2fit}
\begin{array}{|c|c||c|c||c|c|}
\hline
g_0& {\rm Fit}&g_0 &{\rm Fit} &g_0 &{\rm Fit} \\\hline
 6.2 & -0.0061 (g-9.19) &
 6.3 & -0.0060 (g-9.22) &
 6.4 & -0.0059 (g-9.24) \\
 6.5 & -0.0059 (g-9.27) &
 6.6 & -0.0058 (g-9.30) &
 6.7 & -0.0057 (g-9.32) \\
 6.8 & -0.0057 (g-9.35) &
 6.9 & -0.0056 (g-9.38) &
 7.0 & -0.0055 (g-9.40)
 \\\hline
\end{array}~~~~~
\eea
}
As expected, the estimated value of $\bar{g}_{cr}^{(1)}$ increases with $g_0$ approaching $7.2$, and  one concludes from the table that $\bar{g}_{cr}^{(1)}> 9.4$. 
If on the other hand one fits the data to $c_2 (g-\bar{g}_{cr}^{(1)})^2$, one gets
{\smaller 
\bea\la{YQ2fit2}
\begin{array}{|c|c||c|c||c|c|}
\hline
g_0& {\rm Fit}&g_0 &{\rm Fit} &g_0 &{\rm Fit} \\\hline
 6.2 & 0.00061 (g-11.66)^2 &
 6.3 & 0.00061 (g-11.67)^2 &
 6.4 & 0.00061 (g-11.67)^2 \\
 6.5 & 0.00061 (g-11.68)^2 &
 6.6 & 0.00061 (g-11.68)^2&
 6.7 & 0.00060 (g-11.69)^2 \\
 6.8 & 0.00060 (g-11.70)^2 &
 6.9 & 0.00060 (g-11.70)^2 &
 7. & 0.00060 (g-11.71)^2
 \\\hline
\end{array}~~~~~
\eea
}
This fitting is much more stable then the linear one, and gives $\bar{g}_{cr}^{(1)}\sim 11.7$. It is certainly possible that decreasing $Y_2(0,g)$ would slow down for larger values of $g$ resulting in a larger estimate of $\bar{g}_{cr}^{(1)}$.

The plots of $E_{\rm Y_2}$  and $\partial_gE_{\rm Y_2}$ are  shown in Figure 10. The rate of change of $E_{\rm Y_2}$ is still increasing, and one cannot make any reliable prediction about its strong coupling behavior.
\begin{figure}[t]
\begin{center}
\includegraphics*[width=0.45\textwidth]{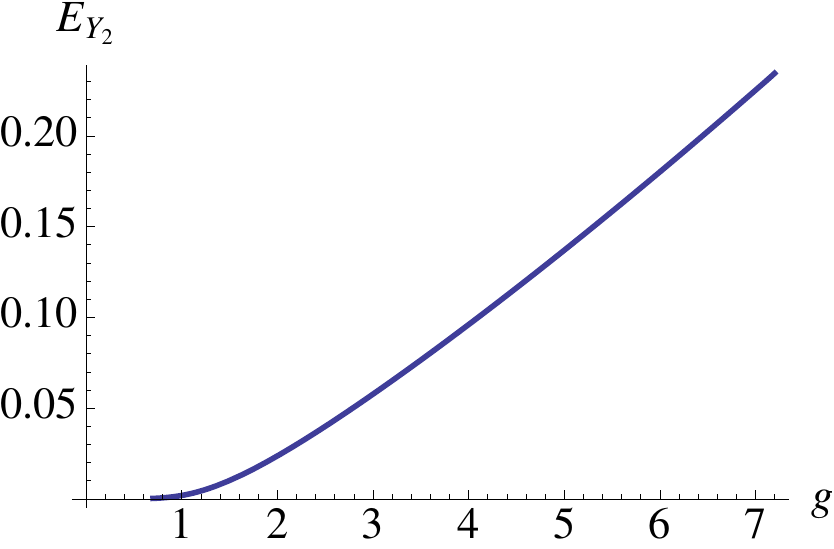}\quad \includegraphics*[width=0.45\textwidth]{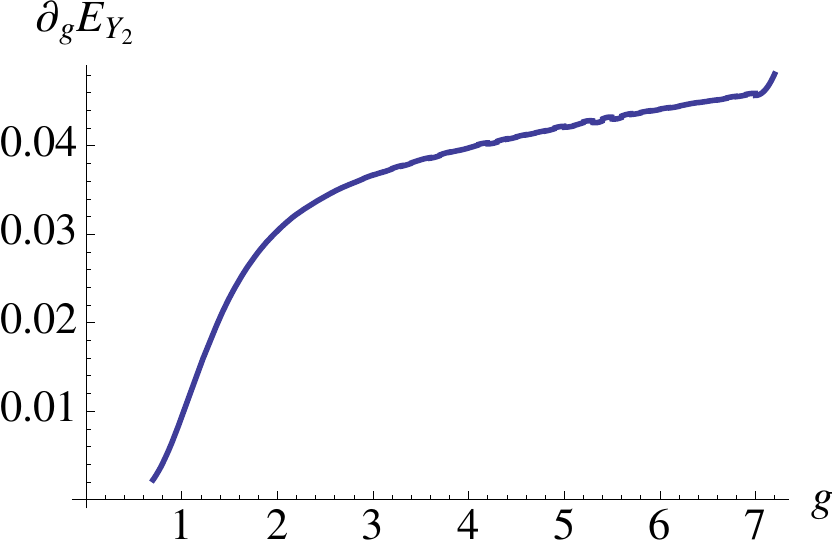}
\end{center}
\caption{\smaller  These are the plots of $E_{\rm Y_2}$ and $\partial_gE_{\rm Y_2}$. 
}
\end{figure}

Figure 10 shows similar plots of $Y_3$-function.  
It is still increasing at $u=0$ but it is clear that it will 
reach its maximum soon.
\begin{figure}[H]
\begin{center}
\includegraphics*[width=0.45\textwidth]{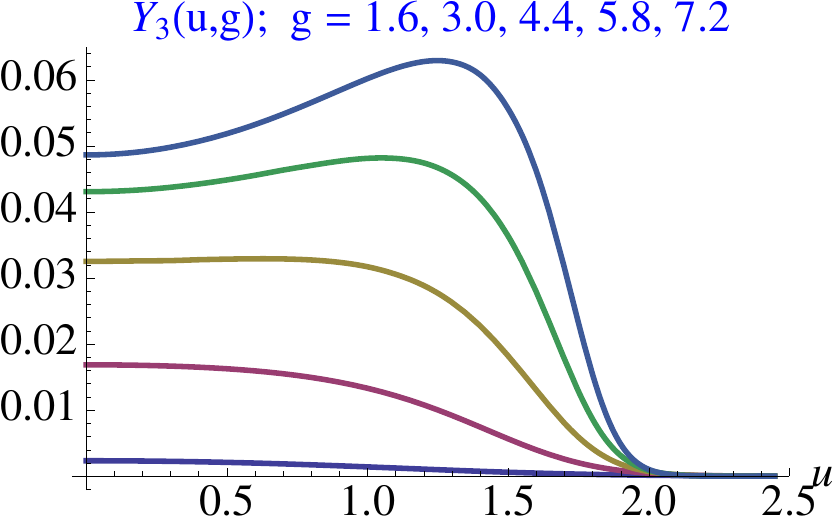}\quad \includegraphics*[width=0.45\textwidth]{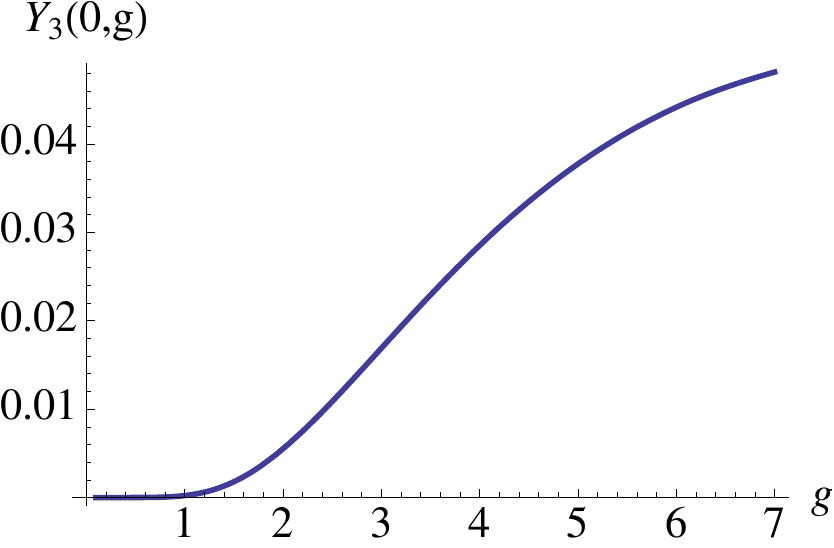}
\end{center}
\caption{\smaller  On the left figure graphs of five $Y_3$-functions with different values of $g$ are shown, and on the right figure the graph of $Y_3(0,g)$ as a function of $g$ is plotted.
}
\end{figure}
Even though $Y_3$ is so small its contribution to the energy is also increasing linearly with $g$, see Figure 12, and moreover the rate of $E_{\rm Y_3}$ has already reached its maximum. 
\begin{figure}[H]
\begin{center}
\includegraphics*[width=0.45\textwidth]{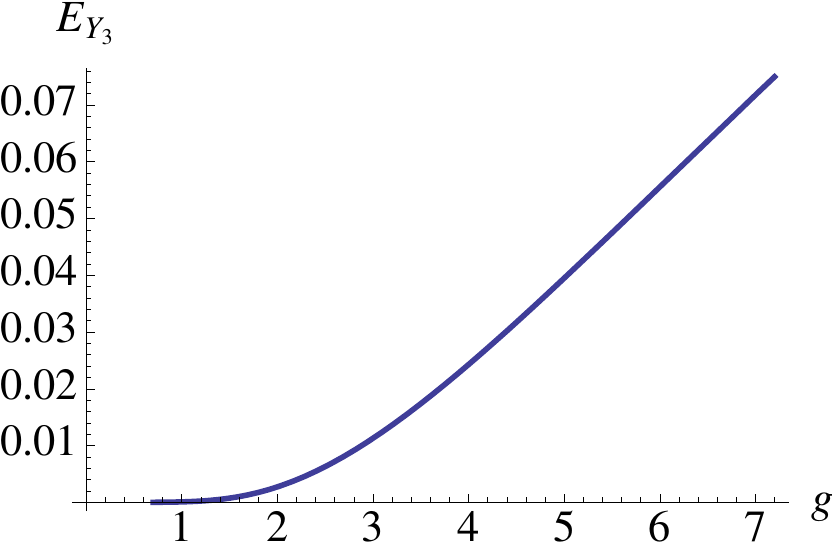}\quad \includegraphics*[width=0.45\textwidth]{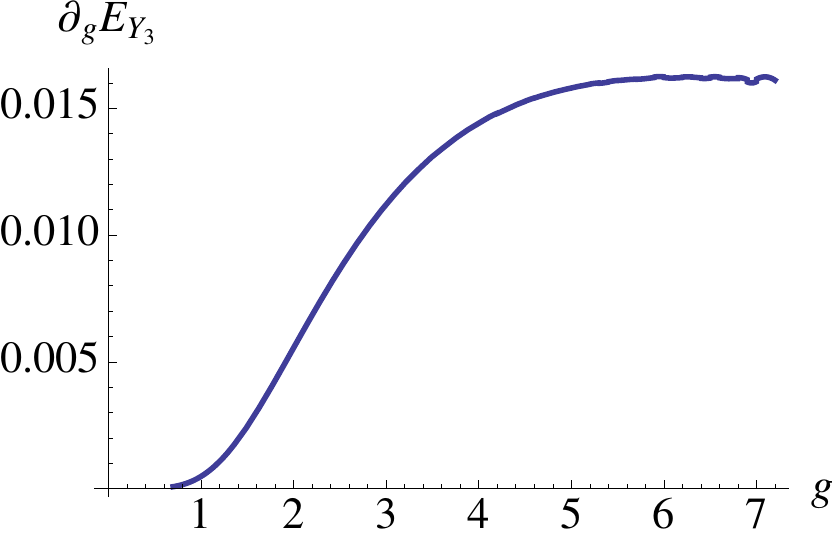}
\end{center}
\caption{\smaller  These are the plots of $E_{\rm Y_3}$ and $\partial_gE_{\rm Y_3}$. 
}
\end{figure}

Next we discuss $Y_{\pm}$-functions, see Figures 13 and 14. 
The first subcritical value can be also determined from $Y_{\pm}$ because their values  vanish at $u=0$, and  
 they are assumed to have the following expansions\footnote{It seems possible that the expansion could be of the form $Y_{\pm}(0,g) \sim (g-\bar{g}_{cr}^{(1)})^3+\cdots\ .$}
$$Y_{\pm}(0,g) \sim g-\bar{g}_{cr}^{(1)}+\cdots\ .$$
\begin{figure}[t]
\begin{center}
\includegraphics*[width=0.45\textwidth]{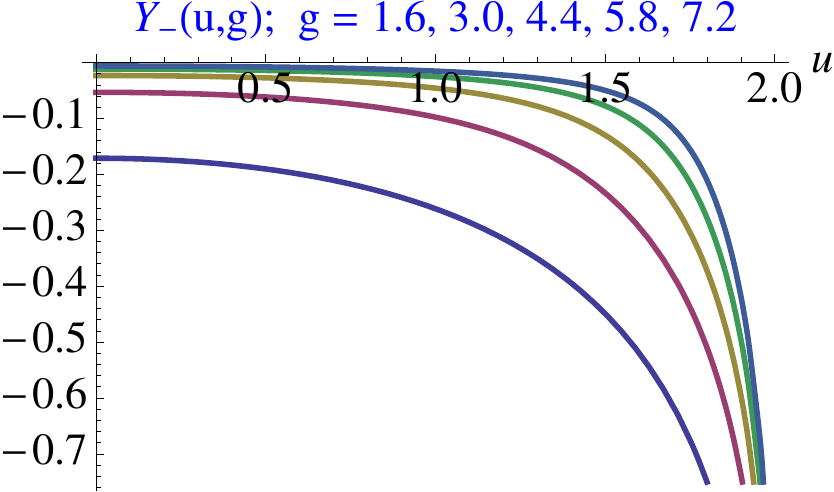}\quad \includegraphics*[width=0.45\textwidth]{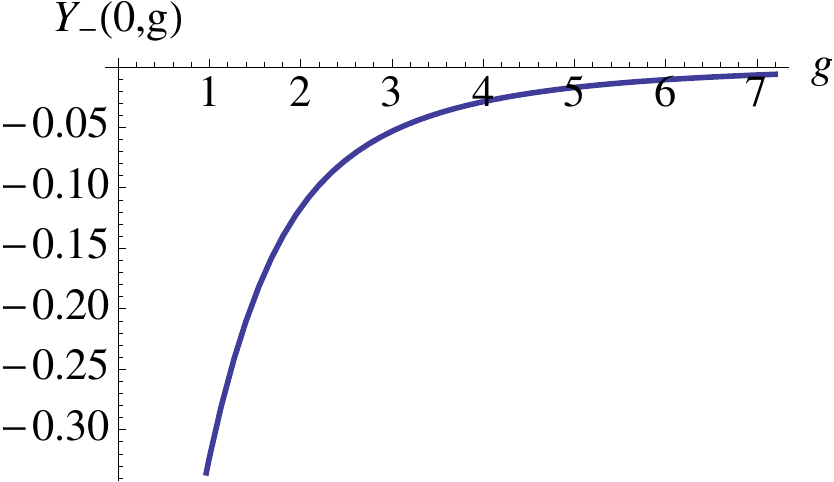}
\end{center}
\caption{\smaller  
$Y_{-}$-functions
}
\end{figure}
\noindent Using again the linear fit, we get the following results for $Y_-(0,g)$
{\smaller \bea\la{Ymfit}
\begin{array}{|c|c||c|c||c|c|}
\hline
g_0& {\rm Fit}&g_0 &{\rm Fit} &g_0 &{\rm Fit} \\\hline
  6.2 & 0.0035 (g-8.93) &
 6.3 & 0.0034 (g-8.98) &
 6.4 & 0.0033 (g-9.03) \\
 6.5 & 0.0033 (g-9.08) &
 6.6 & 0.0032 (g-9.13) &
 6.7 & 0.0031 (g-9.18) \\
 6.8 & 0.0030 (g-9.23) &
 6.9 & 0.0030 (g-9.28) &
 7. & 0.0029 (g-9.33)
 \\\hline
\end{array}~~~
\eea}
\noindent  The results in the table \eqref{Ymfit} are obviously compatible with those in table \eqref{YQ2fit} but the estimated subcritical values $\bar{g}_{cr}^{(1)}$ in  \eqref{Ymfit}  appear to be slightly less than  the  ones from  \eqref{YQ2fit}.
A better estimate of $\bar{g}_{cr}^{(1)}$ is obtained by fitting the data to $c_1 (g-\bar{g}_{cr}^{(1)})+c_3 (g-\bar{g}_{cr}^{(1)})^3$
{\smaller \bea\la{Ymfit2}
\begin{array}{|c|c||c|c|}
\hline
g_0& {\rm Fit}&g_0 &{\rm Fit} \\\hline
 6.2 & 0.000045 (g-11.14)^3+0.00087 (g-11.14) &
 6.3 & 0.000043 (g-11.19)^3+0.00085 (g-11.19) \\
 6.4 & 0.000042 (g-11.25)^3+0.00083 (g-11.25) &
 6.5 & 0.000041 (g-11.30)^3+0.00081 (g-11.30) \\
 6.6 & 0.000040 (g-11.36)^3+0.00079 (g-11.36) &
 6.7 & 0.000039 (g-11.41)^3+0.00078 (g-11.41) \\
 6.8 & 0.000038 (g-11.46)^3+0.00076 (g-11.46) &
 6.9 & 0.000037 (g-11.51)^3+0.00074 (g-11.51) 
 \\\hline
\end{array}~~~
\eea}
with the results similar to those from  \eqref{YQ2fit2}.

Since $Y_{+}(0,g)$ is still pretty far from 0 for the values of $g$ we are dealing with, its linear extrapolation to larger values of $g$ would not give very reliable results. 
\begin{figure}[t]
\begin{center}
\includegraphics*[width=0.45\textwidth]{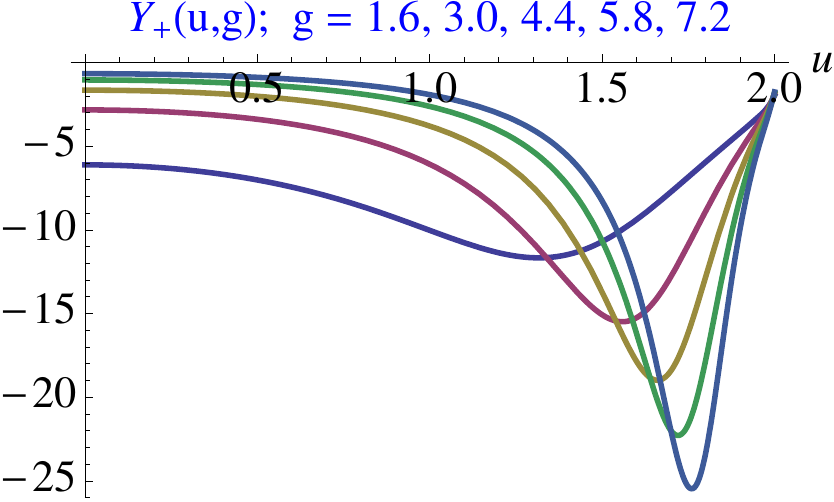}\quad \includegraphics*[width=0.45\textwidth]{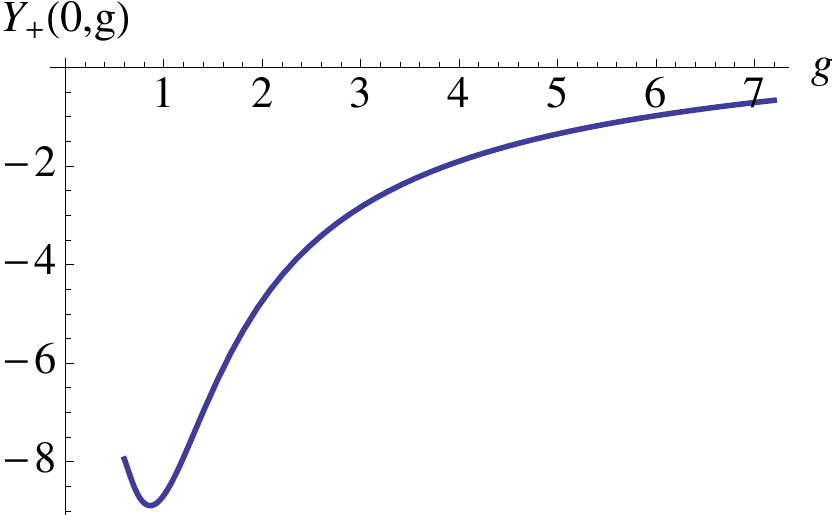}
\end{center}
\caption{\smaller  
$Y_{+}$- and $Y_{-}$-functions  are approaching $0$ at $u=0$ with $g$ increasing. For $g>\bar{g}_{cr}^{(1)}$ they become positive at $u=0$, and asymptote to their ground state value $Y_\pm^{\rm gr\, st}=1$ for very large $g$.}
\end{figure}
\noindent Indeed, 
fitting our data for $Y_{+}(0,g)$
in the interval $[g_0,7.2]$ to the linear function $c_1 (g-\bar{g}_{cr}^{(1)})$, we get the results shown in \eqref{Ypfit}.
{\smaller \bea\la{Ypfit}
\begin{array}{|c|c||c|c||c|c|}
\hline
g_0& {\rm Fit}&g_0 &{\rm Fit} &g_0 &{\rm Fit} \\\hline
 6.2 & 0.25 (g-9.83) &
 6.3 & 0.25 (g-9.88) &
 6.4 & 0.24 (g-9.92) \\
 6.5 & 0.24 (g-9.97) &
 6.6 & 0.24 (g-10.01) &
 6.7 & 0.23 (g-10.05) \\
 6.8 & 0.23 (g-10.10) &
 6.9 & 0.23(g-10.14) &
 7. & 0.22 (g-10.18)
 \\\hline
\end{array}~~~
\eea}
Let us also mention that if one uses $c_3(g-\bar g_{cr}^{(1)})^3$ as a fitting function one gets the largest of all the estimates: $\bar g_{cr}^{(1)}>16$. 

Next we discuss $Y_{1|vw}$-function, see Figure 15.
\begin{figure}[t]
\begin{center}
\includegraphics*[width=0.45\textwidth]{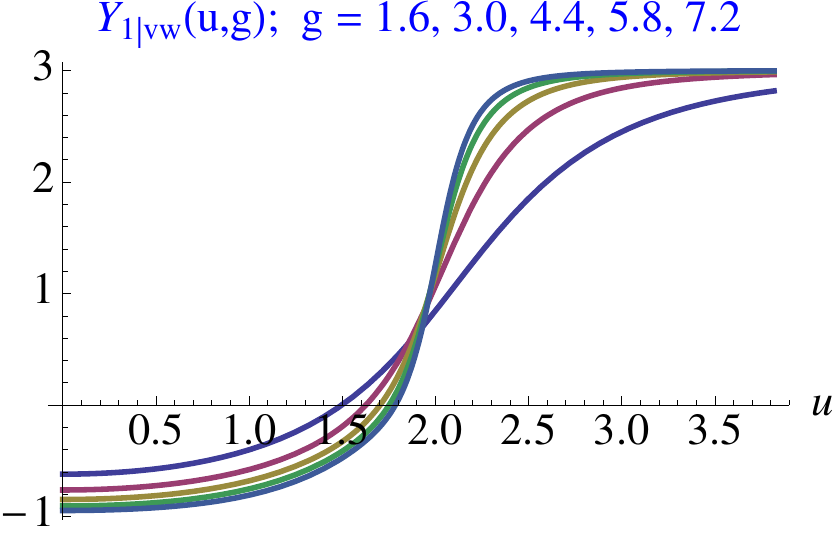}\quad \includegraphics*[width=0.45\textwidth]{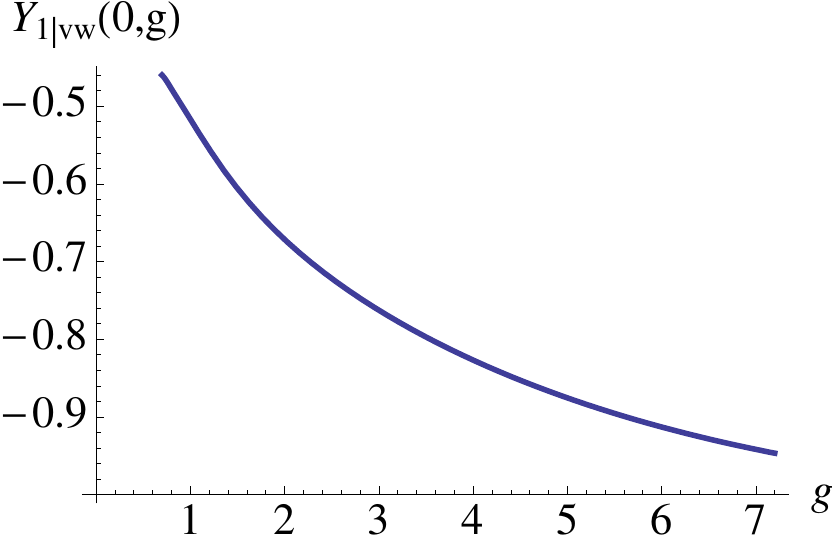}
\end{center}
\caption{\smaller  
$Y_{1|vw}$-function is approaching $-1$ at $u=0$ with $g$ increasing. It also approaches its ground state value at $u=\infty$  faster for larger $g$.
}
\end{figure}
This is the function that determines the first critical value of $g$ because its value at $u=0$ has the following behavior in the vicinity  of $g=g_{cr}^{(1)}$
$$Y_{1|vw}(0,g)+1 \sim (g-{g}_{cr}^{(1)})^2+\cdots\ .$$

Using the quadratic fitting function $c_2 (g-\bar{g}_{cr}^{(1)})^2$, we get the following results
{\smaller \bea\la{Yvw1fit}
\begin{array}{|c|c||c|c||c|c|}
\hline
g_0& {\rm Fit}&g_0 &{\rm Fit} &g_0 &{\rm Fit} \\\hline
  6.2 & 0.0028 (g-11.52)^2 &
 6.3 & 0.0028 (g-11.54)^2 &
 6.4 & 0.0028 (g-11.55)^2 \\
 6.5 & 0.0028 (g-11.57)^2 &
 6.6 & 0.0028 (g-11.59)^2 &
 6.7 & 0.0027 (g-11.60)^2 \\
 6.8 & 0.0027 (g-11.62)^2 &
 6.9 & 0.0027(g-11.64)^2 &
 7. & 0.0027 (g-11.65)^2
 \\\hline
\end{array}~~~
\eea}
The results in the table \eqref{Yvw1fit} are compatible with those in table \eqref{YQ2fit2}. For  all values of $g_0$ the estimated critical value ${g}_{cr}^{(1)}$ appears to be less than  the corresponding subcritical one from  \eqref{YQ2fit2}. They  still do not differ much, and it is what one gets from the analysis of asymptotic Y-functions \cite{AFS09}. 

We conclude from the table  \eqref{Yvw1fit} that ${g}_{cr}^{(1)}> 11.6$, and it seems reasonable to expect that  ${g}_{cr}^{(1)}$ would not exceed $12.0$ (it might appear to be a too optimistic expectation), so the first critical value of $\lam$ would be in the interval $5300<\lam_{cr}^{(1)} <5700$.

The first subcritical value can be also determined from $Y_{2|vw}$, see Figure 16, because its value at $u=0$ vanishes, and  
 it has the following expansion
$$Y_{2|vw}(0,g) \sim g-\bar{g}_{cr}^{(1)}+\cdots\ .$$
\begin{figure}[t]
\begin{center}
\includegraphics*[width=0.45\textwidth]{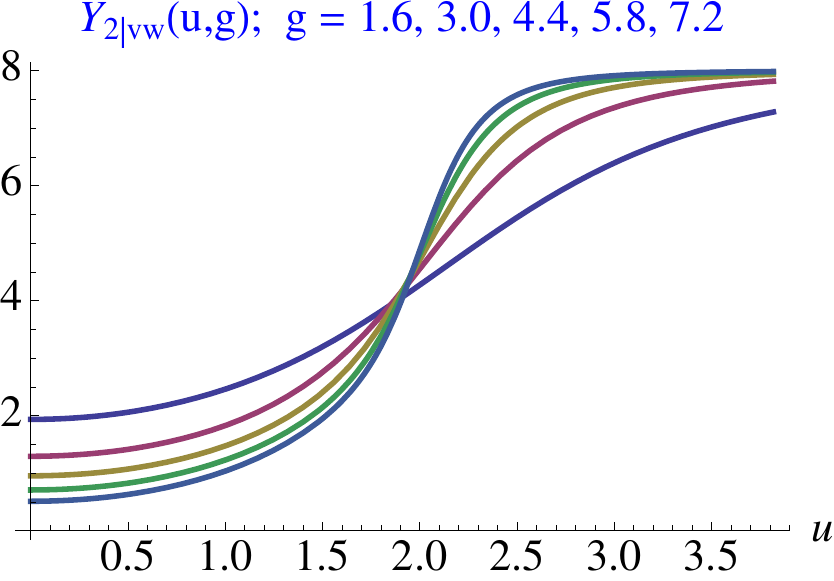}\quad \includegraphics*[width=0.45\textwidth]{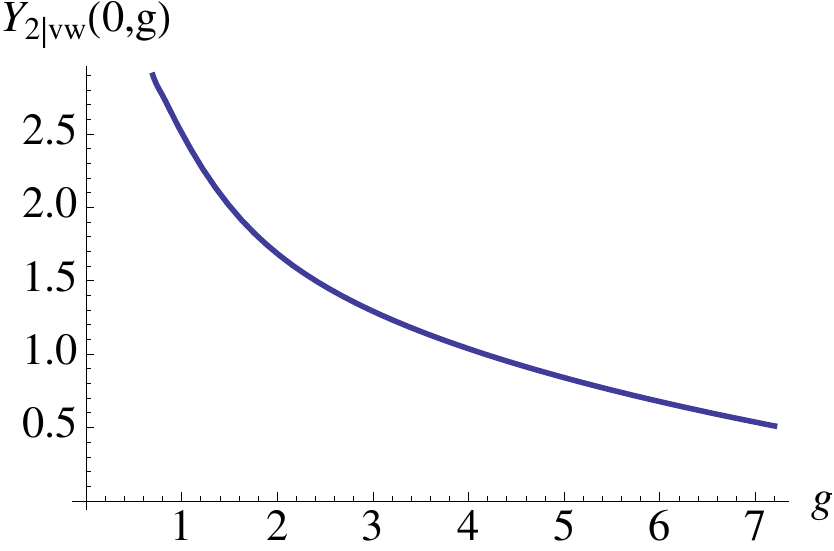}
\end{center}
\caption{\smaller  
$Y_{2|vw}$-function is approaching $0$ at $u=0$ with $g$ increasing. It also approaches its ground state value at $u=\infty$  faster for larger $g$.
}
\end{figure}
$Y_{2|vw}(0,g)$ is also pretty far from 0 and its linear extrapolation to larger values of $g$ gives results slightly lower than those from table \eqref{YQ2fit2}
{\smaller \bea\la{Yvw2fit}
\begin{array}{|c|c||c|c||c|c|}
\hline
g_0& {\rm Fit}&g_0 &{\rm Fit} &g_0 &{\rm Fit} \\\hline
6.2 & -0.137 (g-10.91) &
 6.3 & -0.136 (g-10.94) &
 6.4 & -0.135 (g-10.97) \\
 6.5 & -0.134 (g-11.00) &
 6.6 & -0.133 (g-11.02) &
 6.7 & -0.132 (g-11.05) \\
 6.8 & -0.131 (g-11.08) &
 6.9 & -0.130 (g-11.10) &
 7. & -0.130 (g-11.13)
 \\\hline
\end{array}~~~~
\eea}
The estimates can be made closer to the previous ones if one fits the data to higher order polynomials.

Let us finally mention that $Y_w$-functions do not show any particular $g$-dependence. They just are increasing very fast, see Figure 17.
\begin{figure}[t]
\begin{center}
\includegraphics*[width=0.45\textwidth]{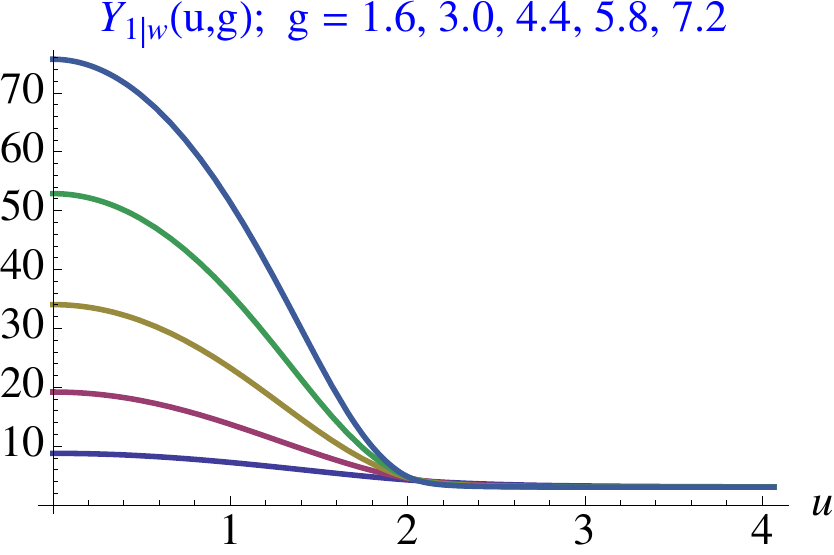}\quad \includegraphics*[width=0.45\textwidth]{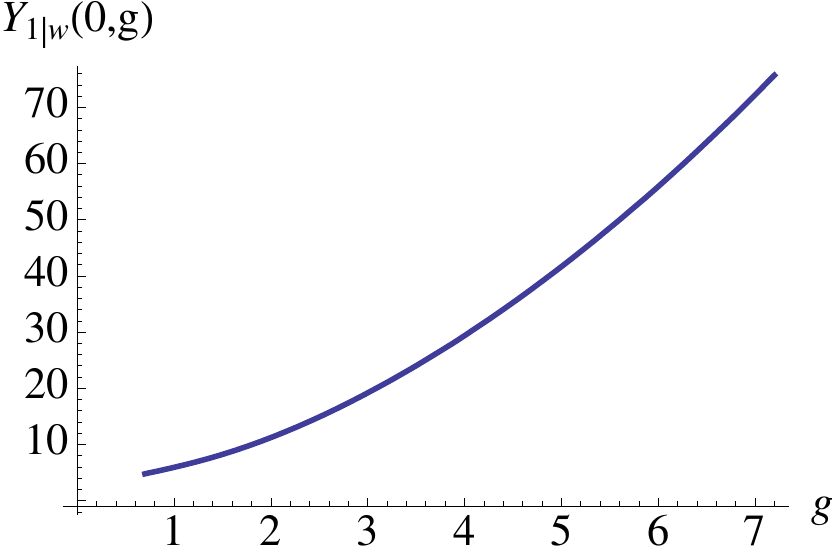}
\end{center}
\caption{\smaller  On the left figure graphs of five $Y_{1|w}$-functions with different values of $g$ are shown, and on the right figure the graph of $Y_{1|w}(0,g)$ as a function of $g$ is plotted.
}
\end{figure}

It is worth stressing that the estimate of $g_{cr}^{(1)}$ was made under an assumption that  the first critical value exists. If it does not then $Y_{1|vw}(0,g)$ would  (or not) approach $-1$ exponentially slow at $g\to\infty$. 

\section{Large $\lam$ expansion from the numerical data}

As was discussed in section 2  at large values of  't Hooft's coupling the Konishi state energy admits an asymptotic expansion in powers of $1/ \sqrt[4]{{\lam}} $
\be
\la{asexp2}
E_K(\lam)=c_{-1} \sqrt[4]{{\lam}}+c_0+\frac{c_1}{
   \sqrt[4]{{\lam}}}+\frac{c_2}{\sqrt{{\lam}}}+\frac{c_3}{\lam^{3/4}}+\frac{c_4}{\lam}+\frac{c_5}{\lam^{5/4}}+ \cdots\,.
   \ee
In this section we try to understand to what extent our numerical data can be used to fix the coefficients $c_i$. We should point out however that in general an asymptotic series cannot be found reliably from numerical data. For example, 
the function $2{1- e^{1000000-\sqrt\lam}\ov 1+e^{1000000-\sqrt\lam}}\sqrt[4]{1+\lam}$ obviously asymptotes to $2 \sqrt[4]{\lam}$ but any numerical computation performed for $\lam <1000000$ would predict that it asymptotes to $-2 \sqrt[4]{\lam}$. Thus, we have to assume first of all that exponentially suppressed terms become very small already at the values of $\lam$ we are dealing with.
Then, a function can approach its asymptotic series monotonically or in oscillations, and it does not seems possible to single out one from numerics.  
In fact, using the standard least-square fitting procedure would always lead to an oscillating behavior of numerical data about a fitting function.  Next, if $\lam$ is not large enough then one may need to make an assumption about the structure of the large $\lam$ expansion, for example to decide if  the series contains all possible terms or some of them vanish. 
Finally, fitting numerical data one should decide how many terms one should keep in an asymptotic series, and what fitting interval one should use. 

Since the precision of our computation is about $10^{-4}$ for $g\sim 7$ it seems reasonable to keep only the terms up to  the $1/\lam^{5/4}$ order in 
 the asymptotic expansion \eqref{asexp2} .
 The fitting is done by using the data with the string tension taking values in the interval $[g_0, g_1]$ where $g_1$ changes from $4.0$ to $7.2$
 (with the step $0.1$), and Mathematica's Fit (or FindFit) functions. 
 The first point of the fitting interval is chosen to be $g_0=1.4$, $\lam\approx 77$ because it is  larger than the inflection point of the Konishi state energy which is $g_{\rm infl}\approx 0.8$, $\lam\approx 25$.

We begin the fitting by making no assumption about the structure of the large $\lam$ expansion.
Below we present the table \eqref{Fitmy0} where we use 
the function in  \eqref{asexp2}  to fit our numerical results in \eqref{Edata}.  
{\smaller \bea\la{Fitmy0}
\begin{array}{|c|c|c|}
\hline
g_1&\lam_1& {\rm Fit} \\\hline
 4. & 632. & 0.0748723-\frac{19.5546}{\lambda ^{5/4}}-\frac{26.8658}{\lambda ^{3/4}}+1.99888
   \sqrt[4]{\lambda }+\frac{0.886704}{\sqrt[4]{\lambda }}+\frac{7.23013}{\sqrt{\lambda
   }}+\frac{40.4154}{\lambda } \\
 4.4 & 764. & 1.3609-\frac{176.565}{\lambda ^{5/4}}-\frac{198.245}{\lambda ^{3/4}}+1.94263
   \sqrt[4]{\lambda }-\frac{11.2876}{\sqrt[4]{\lambda }}+\frac{68.3193}{\sqrt{\lambda
   }}+\frac{295.295}{\lambda } \\
 4.8 & 910. & 0.94197-\frac{117.655}{\lambda ^{5/4}}-\frac{137.196}{\lambda ^{3/4}}+1.96036
   \sqrt[4]{\lambda }-\frac{7.19535}{\sqrt[4]{\lambda }}+\frac{47.1638}{\sqrt{\lambda
   }}+\frac{202.045}{\lambda } \\
 5.2 & 1070. & 0.443739-\frac{41.3538}{\lambda ^{5/4}}-\frac{60.5348}{\lambda ^{3/4}}+1.98102
   \sqrt[4]{\lambda }-\frac{2.23404}{\sqrt[4]{\lambda }}+\frac{21.0468}{\sqrt{\lambda
   }}+\frac{83.0606}{\lambda } \\
 5.6 & 1240. & -0.0777338+\frac{45.6551}{\lambda ^{5/4}}+\frac{24.2251}{\lambda ^{3/4}}+2.0022
   \sqrt[4]{\lambda }+\frac{3.06072}{\sqrt[4]{\lambda }}-\frac{7.33942}{\sqrt{\lambda
   }}-\frac{50.6135}{\lambda } \\
 6. & 1420. & -0.439602+\frac{110.23}{\lambda ^{5/4}}+\frac{85.6348}{\lambda ^{3/4}}+2.01667
   \sqrt[4]{\lambda }+\frac{6.79178}{\sqrt[4]{\lambda }}-\frac{27.6317}{\sqrt{\lambda
   }}-\frac{148.678}{\lambda } \\
 6.4 & 1620. & -0.131008+\frac{51.482}{\lambda ^{5/4}}+\frac{31.0414}{\lambda ^{3/4}}+2.00453
   \sqrt[4]{\lambda }+\frac{3.56246}{\sqrt[4]{\lambda }}-\frac{9.8231}{\sqrt{\lambda
   }}-\frac{60.4474}{\lambda } \\
 6.8 & 1830. & 0.201473-\frac{15.5603}{\lambda ^{5/4}}-\frac{30.0019}{\lambda ^{3/4}}+1.99163
   \sqrt[4]{\lambda }+\frac{0.0366606}{\sqrt[4]{\lambda }}+\frac{9.86239}{\sqrt{\lambda
   }}+\frac{39.2596}{\lambda } \\
 7.2 & 2050. & 0.210769-\frac{17.5595}{\lambda ^{5/4}}-\frac{31.7814}{\lambda ^{3/4}}+1.99128
   \sqrt[4]{\lambda }-\frac{0.0634158}{\sqrt[4]{\lambda }}+\frac{10.429}{\sqrt{\lambda
   }}+\frac{42.2007}{\lambda }
   \\\hline
   \end{array}~~~~
\eea}
To better visualize the results in Figure 18 we also plot the coefficients $c_{-1}$ and $c_0$.
\begin{figure}[t]
\begin{center}
\includegraphics*[width=0.45\textwidth]{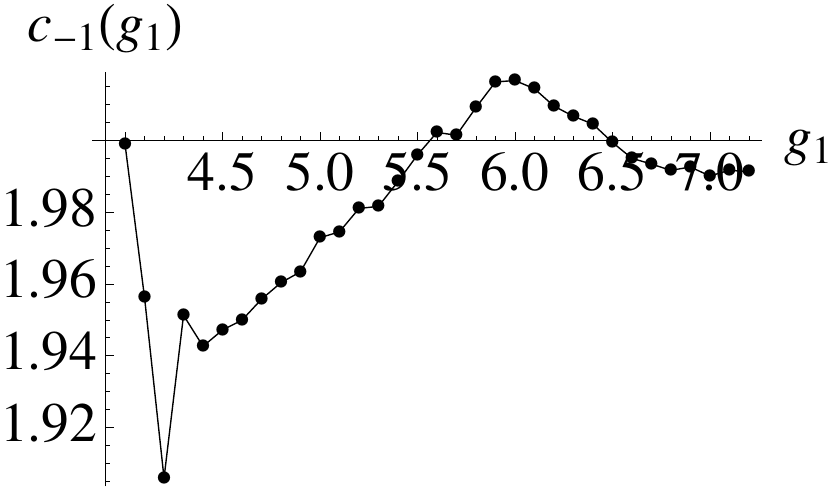}\quad \includegraphics*[width=0.45\textwidth]{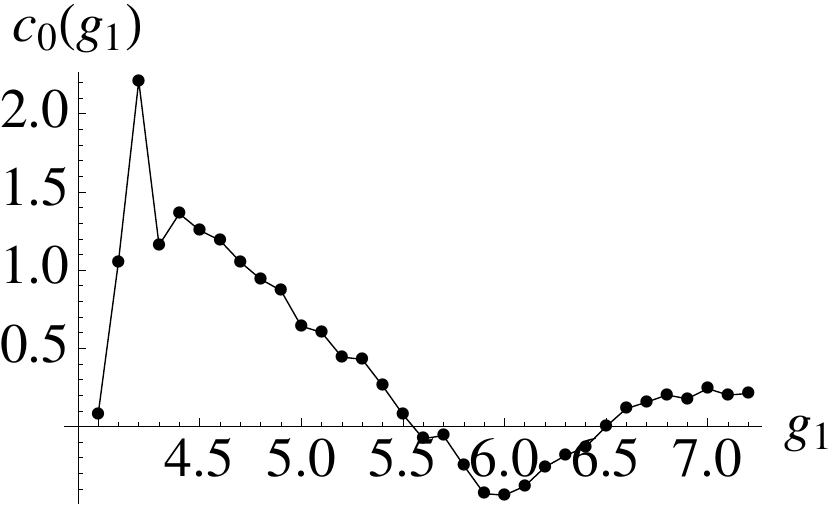}
\end{center}
\caption{\smaller  
On the left and right figures the graphs of $c_{-1}$ and $c_0$ as functions of $g_1$ are plotted. The fitting is done  without any constraint imposed on $c_i$.
}
\end{figure}
As one can see, the coefficient $c_{-1}$ of the leading term is oscillating about 2, and for $g_1>6.2$ it  is getting quite close to 2. Its average value is 1.983.  The coefficient $c_0$ is oscillating about 0 but it is not really small with the average equal to $0.385$. There is however an obvious correlation between the values of $c_{-1}$ and $c_0$ -- the closer $c_{-1}$ is to 2 the closer $c_0$ is to 0.

The subleading coefficients are however not fixed at all and take very different values depending on the fitting interval used. 
The fitting therefore is not stable, and the fitting function strongly depends on the fitting interval. One may conclude that the numerical data allows one to fix only the leading coefficient in the strong coupling expansion. One should remember however that the series \eqref{asexp2} is only asymptotic, and fixing its coefficients by using the data for these not very large values of $\lam$ is not that straightforward as it is for a convergent series.  

To proceed let us fix the leading coefficient to be 2. Then, fitting the numerical data, one finds the results in table \eqref{Fitmy1} and Figure 19.
{\smaller \bea\la{Fitmy1}
\begin{array}{|c|c|c|}
\hline
g_1&\lam_1& {\rm Fit} \\\hline
 4. & 632. & 0.0487754-\frac{16.0827}{\lambda ^{5/4}}-\frac{23.1977}{\lambda ^{3/4}}+2
   \sqrt[4]{\lambda }+\frac{1.13832}{\sqrt[4]{\lambda }}+\frac{5.94492}{\sqrt{\lambda
   }}+\frac{34.8688}{\lambda } \\
 4.4 & 764. & -0.00945538+\frac{26.2666}{\lambda ^{5/4}}+\frac{7.84673}{\lambda ^{3/4}}+2
   \sqrt[4]{\lambda }+\frac{2.2426}{\sqrt[4]{\lambda }}-\frac{2.36707}{\sqrt{\lambda
   }}-\frac{22.6721}{\lambda } \\
 4.8 & 910. & -0.0269174+\frac{40.3386}{\lambda ^{5/4}}+\frac{17.7895}{\lambda ^{3/4}}+2
   \sqrt[4]{\lambda }+\frac{2.58159}{\sqrt[4]{\lambda }}-\frac{4.97525}{\sqrt{\lambda
   }}-\frac{41.4559}{\lambda } \\
 5.2 & 1070. & -0.030125+\frac{43.0799}{\lambda ^{5/4}}+\frac{19.6865}{\lambda ^{3/4}}+2
   \sqrt[4]{\lambda }+\frac{2.64471}{\sqrt[4]{\lambda }}-\frac{5.46707}{\sqrt{\lambda
   }}-\frac{45.0787}{\lambda } \\
 5.6 & 1240. & -0.0216031+\frac{34.8027}{\lambda ^{5/4}}+\frac{14.2113}{\lambda ^{3/4}}+2
   \sqrt[4]{\lambda }+\frac{2.47195}{\sqrt[4]{\lambda }}-\frac{4.08328}{\sqrt{\lambda
   }}-\frac{34.3737}{\lambda } \\
 6. & 1420. & -0.00723624+\frac{20.0648}{\lambda ^{5/4}}+\frac{4.64678}{\lambda ^{3/4}}+2
   \sqrt[4]{\lambda }+\frac{2.17693}{\sqrt[4]{\lambda }}-\frac{1.69214}{\sqrt{\lambda
   }}-\frac{15.4861}{\lambda } \\
 6.4 & 1620. & -0.0115075+\frac{24.7422}{\lambda ^{5/4}}+\frac{7.61427}{\lambda ^{3/4}}+2
   \sqrt[4]{\lambda }+\frac{2.266}{\sqrt[4]{\lambda }}-\frac{2.42443}{\sqrt{\lambda
   }}-\frac{21.4161}{\lambda } \\
 6.8 & 1830. & -0.0230903+\frac{38.1119}{\lambda ^{5/4}}+\frac{15.9437}{\lambda ^{3/4}}+2
   \sqrt[4]{\lambda }+\frac{2.5106}{\sqrt[4]{\lambda }}-\frac{4.45848}{\sqrt{\lambda
   }}-\frac{38.2204}{\lambda } \\
 7.2 & 2050. & -0.0269859+\frac{42.8933}{\lambda ^{5/4}}+\frac{18.8607}{\lambda ^{3/4}}+2
   \sqrt[4]{\lambda }+\frac{2.59408}{\sqrt[4]{\lambda }}-\frac{5.16215}{\sqrt{\lambda
   }}-\frac{44.1706}{\lambda }
    \\\hline
   \end{array}~~~~
\eea}
\begin{figure}[t]
\begin{center}
\includegraphics*[width=0.45\textwidth]{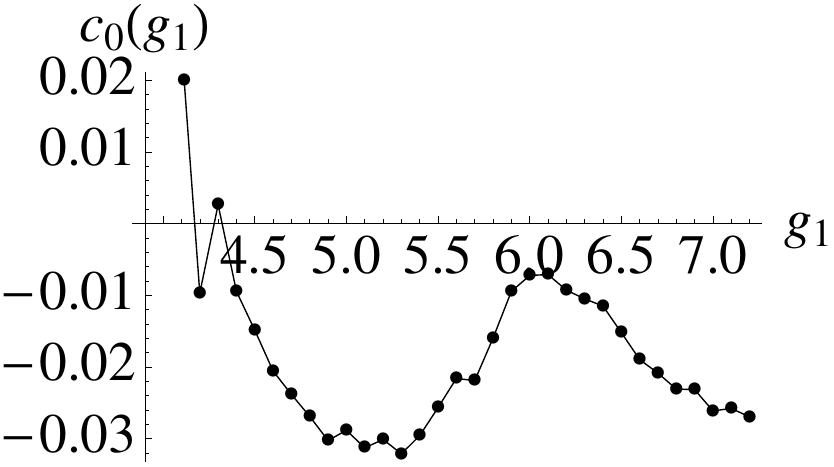}\quad \includegraphics*[width=0.45\textwidth]{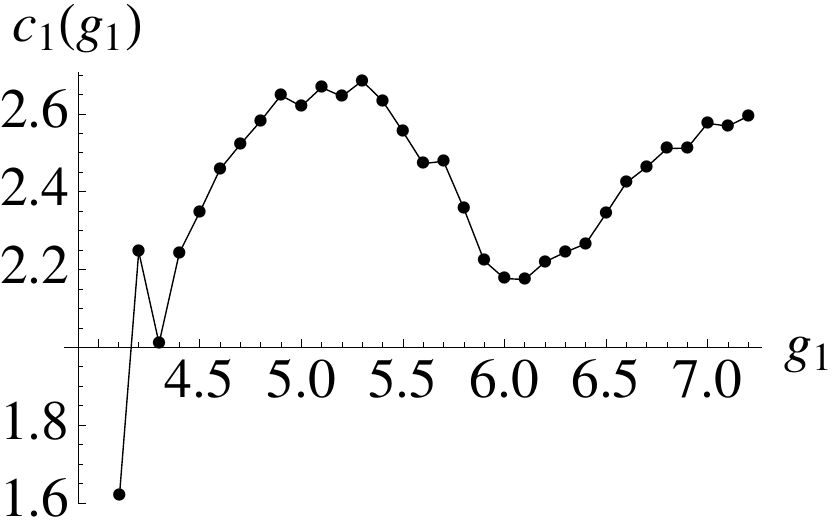}
\end{center}
\caption{\smaller  On the left and right figures the graphs of $c_{0}$ and $c_1$ as functions of $g_1$ are plotted. The fitting is done with $c_{-1}=2$.
}
\end{figure}
The coefficient $c_0$ is  now much smaller but it stops oscillating about 0. The average value of $c_0$ is  $-0.016$, and $c_0$ is decreasing with $g_1$ increasing. 
The subleading coefficient $c_1$ is not close to 2, it has an average 2.37, and is greater than 2 for almost all values of $g_1$. 
There is still a correlation between the values of $c_{0}$ and $c_1$ -- the closer $c_{0}$ is to 0 the closer $c_1$ is to 2.  One sees that for the range of $\lambda$ we are analyzing the contribution of the constant term is much smaller than the one of the subleading term $c_1/\sqrt[4]\lam$ and, therefore, it is reasonable to assume that $c_0=0$.  

Fixing now the leading and constant coefficients to be 2 and 0, respectively, one finds the results in table \eqref{Fitmy2} and Figure 20.
{\smaller \bea\la{Fitmy2}
\begin{array}{|c|c|c|}
\hline
g_1&\lam_1& {\rm Fit} \\\hline
 4. & 632. & \frac{22.2908}{\lambda ^{5/4}}+\frac{4.1118}{\lambda ^{3/4}}+2 \sqrt[4]{\lambda
   }+\frac{2.07902}{\sqrt[4]{\lambda }}-\frac{1.25263}{\sqrt{\lambda }}-\frac{16.5214}{\lambda }
   \\
 4.4 & 764. & \frac{18.0024}{\lambda ^{5/4}}+\frac{2.18271}{\lambda ^{3/4}}+2 \sqrt[4]{\lambda
   }+\frac{2.0558}{\sqrt[4]{\lambda }}-\frac{0.905321}{\sqrt{\lambda }}-\frac{11.8034}{\lambda }
   \\
 4.8 & 910. & \frac{14.458}{\lambda ^{5/4}}+\frac{0.641502}{\lambda ^{3/4}}+2 \sqrt[4]{\lambda
   }+\frac{2.03799}{\sqrt[4]{\lambda }}-\frac{0.633195}{\sqrt{\lambda }}-\frac{7.96653}{\lambda
   } \\
 5.2 & 1070. & \frac{11.4777}{\lambda ^{5/4}}-\frac{0.621309}{\lambda ^{3/4}}+2 \sqrt[4]{\lambda
   }+\frac{2.02384}{\sqrt[4]{\lambda }}-\frac{0.413548}{\sqrt{\lambda }}-\frac{4.77967}{\lambda
   } \\
 5.6 & 1240. & \frac{10.2483}{\lambda ^{5/4}}-\frac{1.13316}{\lambda ^{3/4}}+2 \sqrt[4]{\lambda
   }+\frac{2.01824}{\sqrt[4]{\lambda }}-\frac{0.325432}{\sqrt{\lambda }}-\frac{3.47594}{\lambda
   } \\
 6. & 1420. & \frac{11.2064}{\lambda ^{5/4}}-\frac{0.748872}{\lambda ^{3/4}}+2 \sqrt[4]{\lambda
   }+\frac{2.02225}{\sqrt[4]{\lambda }}-\frac{0.390144}{\sqrt{\lambda }}-\frac{4.47424}{\lambda
   } \\
 6.4 & 1620. & \frac{9.64851}{\lambda ^{5/4}}-\frac{1.36445}{\lambda ^{3/4}}+2 \sqrt[4]{\lambda
   }+\frac{2.01594}{\sqrt[4]{\lambda }}-\frac{0.287401}{\sqrt{\lambda }}-\frac{2.86239}{\lambda
   } \\
 6.8 & 1830. & \frac{5.80801}{\lambda ^{5/4}}-\frac{2.85568}{\lambda ^{3/4}}+2 \sqrt[4]{\lambda
   }+\frac{2.00097}{\sqrt[4]{\lambda }}-\frac{0.041104}{\sqrt{\lambda }}+\frac{1.07875}{\lambda
   } \\
 7.2 & 2050. & \frac{2.78511}{\lambda ^{5/4}}-\frac{4.00844}{\lambda ^{3/4}}+2 \sqrt[4]{\lambda
   }+\frac{1.98967}{\sqrt[4]{\lambda }}+\frac{0.147219}{\sqrt{\lambda }}+\frac{4.15477}{\lambda
   }
\\\hline   
\end{array}~~~~
\eea}
\begin{figure}[t]
\begin{center}
\includegraphics*[width=0.45\textwidth]{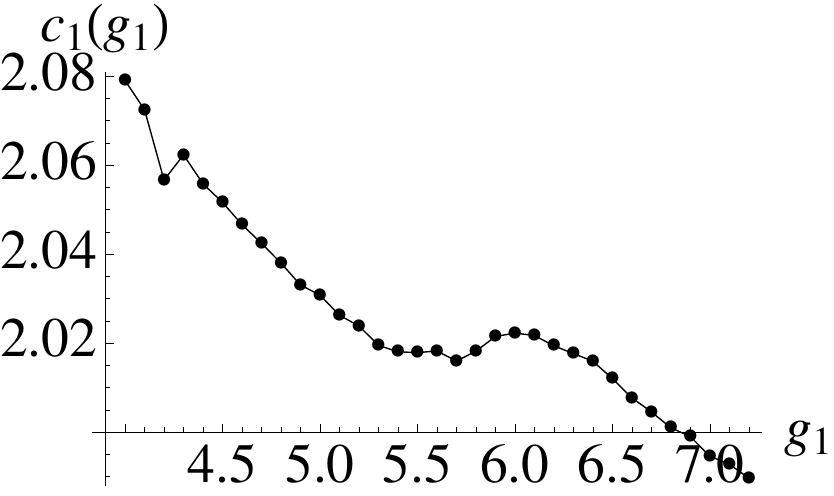}\quad \includegraphics*[width=0.45\textwidth]{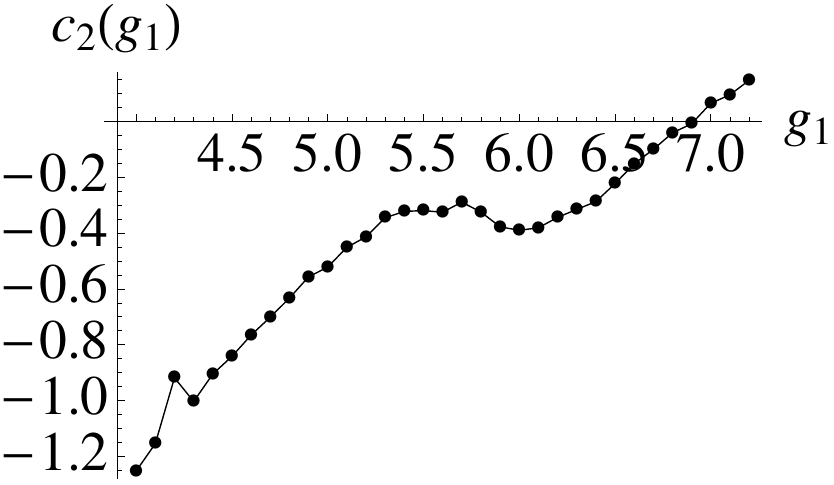}
\end{center}
\caption{\smaller  On the left and right figures the graphs of $c_{1}$ and $c_2$ as functions of $g_1$ are plotted. The fitting is done with $c_{-1}=2$ and $c_0=0$.
}
\end{figure}
It is clear from the table and Figure 20 that fixing $c_{-1}=2$ and $c_0=0$ makes the first nontrivial subleading coefficient $c_1$ to be very close to 2. Its average value is $\approx 2.026$. The next coefficient $c_2$  still varies significantly and its average is $-0.435$. However $c_2$ is  increasing and becomes very close to 0 for $g_1=6.9$.

Since the coefficient $c_1$ is so close to 2, let us proceed by fixing $c_{-1}=2, c_0=0$, and $c_1=2$. Then, one obtains the following table and Figure 21
{\smaller \bea\la{Fitmy}
\begin{array}{|c|c|c|}
\hline
g_1&\lam_1& {\rm Fit} \\\hline
 4. & 632. & \frac{6.48139}{\text{la}^{5/4}}-\frac{2.7364}{\text{la}^{3/4}}+2
   \sqrt[4]{\text{la}}+\frac{2}{\sqrt[4]{\text{la}}}-\frac{0.044888}{\sqrt{\text{la}}}+\frac{0.5
   56775}{\text{la}} \\
 4.4 & 764. & \frac{5.92711}{\text{la}^{5/4}}-\frac{2.86533}{\text{la}^{3/4}}+2
   \sqrt[4]{\text{la}}+\frac{2}{\sqrt[4]{\text{la}}}-\frac{0.0331335}{\sqrt{\text{la}}}+\frac{1.
   02244}{\text{la}} \\
 4.8 & 910. & \frac{5.63528}{\text{la}^{5/4}}-\frac{2.93171}{\text{la}^{3/4}}+2
   \sqrt[4]{\text{la}}+\frac{2}{\sqrt[4]{\text{la}}}-\frac{0.0271635}{\sqrt{\text{la}}}+\frac{1.
   26505}{\text{la}} \\
 5.2 & 1070. & \frac{5.57229}{\text{la}^{5/4}}-\frac{2.94569}{\text{la}^{3/4}}+2
   \sqrt[4]{\text{la}}+\frac{2}{\sqrt[4]{\text{la}}}-\frac{0.0259249}{\sqrt{\text{la}}}+\frac{1.
   31682}{\text{la}} \\
 5.6 & 1240. & \frac{5.45878}{\text{la}^{5/4}}-\frac{2.97011}{\text{la}^{3/4}}+2
   \sqrt[4]{\text{la}}+\frac{2}{\sqrt[4]{\text{la}}}-\frac{0.0238023}{\sqrt{\text{la}}}+\frac{1.
   40878}{\text{la}} \\
 6. & 1420. & \frac{5.03661}{\text{la}^{5/4}}-\frac{3.0599}{\text{la}^{3/4}}+2
   \sqrt[4]{\text{la}}+\frac{2}{\sqrt[4]{\text{la}}}-\frac{0.0160542}{\sqrt{\text{la}}}+\frac{1.
   749}{\text{la}} \\
 6.4 & 1620. & \frac{5.00365}{\text{la}^{5/4}}-\frac{3.06689}{\text{la}^{3/4}}+2
   \sqrt[4]{\text{la}}+\frac{2}{\sqrt[4]{\text{la}}}-\frac{0.0154523}{\sqrt{\text{la}}}+\frac{1.
   77553}{\text{la}} \\
 6.8 & 1830. & \frac{5.51203}{\text{la}^{5/4}}-\frac{2.96201}{\text{la}^{3/4}}+2
   \sqrt[4]{\text{la}}+\frac{2}{\sqrt[4]{\text{la}}}-\frac{0.0243275}{\sqrt{\text{la}}}+\frac{1.
   37154}{\text{la}} \\
 7.2 & 2050. & \frac{6.08044}{\text{la}^{5/4}}-\frac{2.84643}{\text{la}^{3/4}}+2
   \sqrt[4]{\text{la}}+\frac{2}{\sqrt[4]{\text{la}}}-\frac{0.0340167}{\sqrt{\text{la}}}+\frac{0.
   922834}{\text{la}}
\\\hline   
\end{array}~~~~
\eea}
\begin{figure}[t]
\begin{center}
\includegraphics*[width=0.45\textwidth]{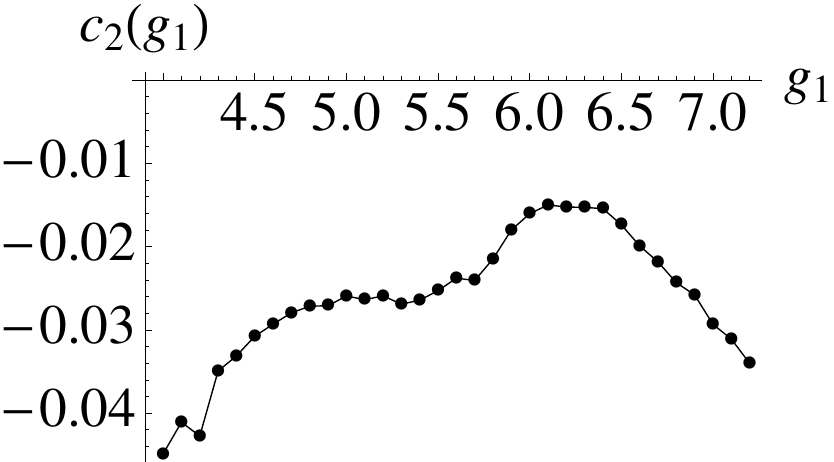}\quad \includegraphics*[width=0.45\textwidth]{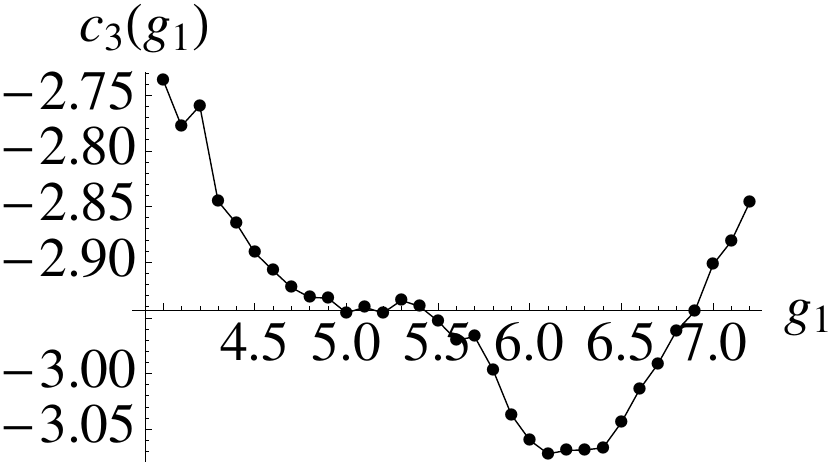}
\end{center}
\caption{\smaller  On the left and right figures the graphs of $c_{2}$ and $c_3$ as functions of $g_1$ are plotted. The fitting is done with $c_{-1}=2$, $c_0=0$ and $c_1=2$.
}
\end{figure}
The fitting with the subleading coefficient equal to 2
makes the coefficient $c_2$ to be very small with the average $-0.026$.
Its contribution to the energy is much smaller than the contribution of the next term $c_3/\lam^{3/4}$, and it is reasonable to assume that $c_2$ is equal to 0.
So, let us proceed by fixing $c_{-1}=2, c_0=0, c_1=2$, and $c_2=0$. Then, one obtains the following table and Figure 22
{\smaller \bea\la{Fitmyy}
\begin{array}{|c|c|c|}
\hline
g_1&\lam_1& {\rm Fit} \\\hline
  4. & 632. & \frac{4.19637}{\text{la}^{5/4}}-\frac{3.24272}{\text{la}^{3/4}}+2
   \sqrt[4]{\text{la}}+\frac{2}{\sqrt[4]{\text{la}}}+\frac{2.43269}{\text{la}} \\
 4.4 & 764. & \frac{4.15138}{\text{la}^{5/4}}-\frac{3.24647}{\text{la}^{3/4}}+2
   \sqrt[4]{\text{la}}+\frac{2}{\sqrt[4]{\text{la}}}+\frac{2.45884}{\text{la}} \\
 4.8 & 910. & \frac{4.11145}{\text{la}^{5/4}}-\frac{3.24973}{\text{la}^{3/4}}+2
   \sqrt[4]{\text{la}}+\frac{2}{\sqrt[4]{\text{la}}}+\frac{2.48182}{\text{la}} \\
 5.2 & 1070. & \frac{4.05719}{\text{la}^{5/4}}-\frac{3.25408}{\text{la}^{3/4}}+2
   \sqrt[4]{\text{la}}+\frac{2}{\sqrt[4]{\text{la}}}+\frac{2.51283}{\text{la}} \\
 5.6 & 1240. & \frac{4.0154}{\text{la}^{5/4}}-\frac{3.2574}{\text{la}^{3/4}}+2
   \sqrt[4]{\text{la}}+\frac{2}{\sqrt[4]{\text{la}}}+\frac{2.5366}{\text{la}} \\
 6. & 1420. & \frac{4.02985}{\text{la}^{5/4}}-\frac{3.25628}{\text{la}^{3/4}}+2
   \sqrt[4]{\text{la}}+\frac{2}{\sqrt[4]{\text{la}}}+\frac{2.52846}{\text{la}} \\
 6.4 & 1620. & \frac{4.00445}{\text{la}^{5/4}}-\frac{3.25825}{\text{la}^{3/4}}+2
   \sqrt[4]{\text{la}}+\frac{2}{\sqrt[4]{\text{la}}}+\frac{2.54273}{\text{la}} \\
 6.8 & 1830. & \frac{3.89399}{\text{la}^{5/4}}-\frac{3.26672}{\text{la}^{3/4}}+2
   \sqrt[4]{\text{la}}+\frac{2}{\sqrt[4]{\text{la}}}+\frac{2.60457}{\text{la}} \\
 7.2 & 2050. & \frac{3.75839}{\text{la}^{5/4}}-\frac{3.27702}{\text{la}^{3/4}}+2
   \sqrt[4]{\text{la}}+\frac{2}{\sqrt[4]{\text{la}}}+\frac{2.68017}{\text{la}}
\\\hline   
\end{array}~~~~
\eea}
\begin{figure}[t]
\begin{center}
\includegraphics*[width=0.45\textwidth]{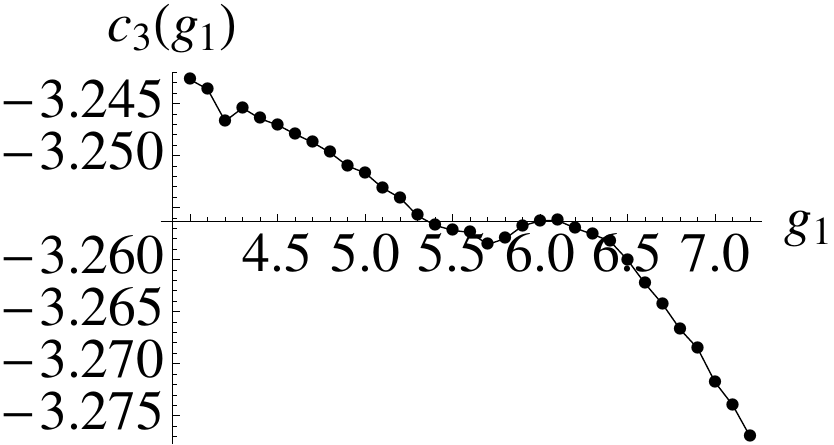}\quad \includegraphics*[width=0.45\textwidth]{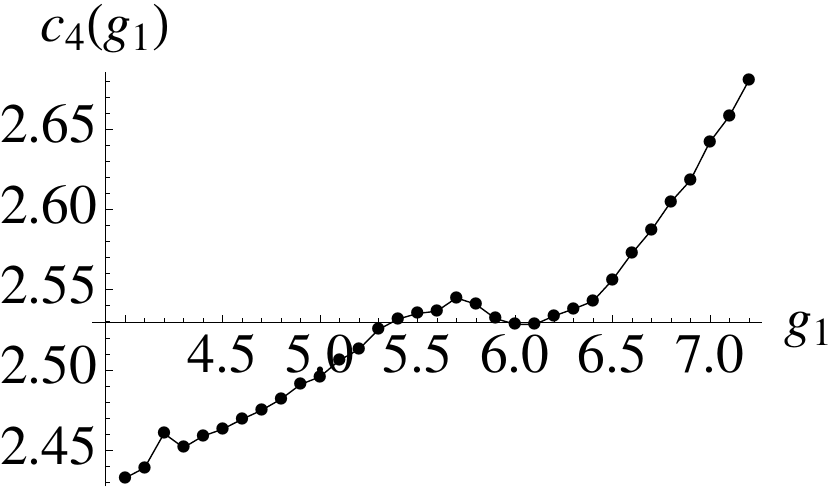}
\end{center}
\caption{\smaller  On the left and right figures the graphs of $c_{2}$ and $c_3$ as functions of $g_1$ are plotted. The fitting is done with $c_{-1}=2$, $c_0=0$ and $c_1=2$.
}
\end{figure}
We see an interesting effect of this fitting. It appears to be rather stable. 
     As one can see from  the table \eqref{Fitmy}, the remaining three coefficients $c_3$, $c_4$ and $c_5$ are not in fact too sensitive to the choice of $g_1$ anymore.
     
To continue we use the average value of $c_3$, and fit the coefficients $c_4$ and $c_5$. Then, we use the average value of $c_4$, and fit the coefficient $c_5$. Finally, taking the average value of $c_5$, we find the following fitting function 
\bea
\la{FFit14}
g_0=1.4:\qquad {\overline E}_K(\lam)= 2
   \sqrt[4]{\lam}+\frac{2}{\sqrt[4]{\lam}}-\frac{3.26}{\lam^{3/4}}+\frac{2.53}{\lam}+
 \frac{4.03}{\lam^{5/4}}\,.~~~~~
 \eea
 Obviously, the fitting function is different from \eqref{EGKV}, in particular the coefficient $c_4$ is not small and gives a significant contribution to the energy. The coefficients $c_i$ in \eqref{FFit14}  depend on the choice of $g_0$.  In the table below we present fitting functions for $1.2\le g_0\le 1.6$
\bea\la{Fitmyy}
\begin{array}{|c|c|}
\hline
g_0& {\overline E}_K\\\hline
 1.2 & 2
   \sqrt[4]{\lam}+\frac{2}{\sqrt[4]{\lam}}-\frac{3.18}{\lam^{3/4}}+\frac{1.89}{\lam}+
 \frac{5.27}{\lam^{5/4}}\\
 1.3 & 2
   \sqrt[4]{\lam}+\frac{2}{\sqrt[4]{\lam}}-\frac{3.22}{\lam^{3/4}}+\frac{2.25}{\lam}+
 \frac{4.57}{\lam^{5/4}}\\
 1.4 & 2
   \sqrt[4]{\lam}+\frac{2}{\sqrt[4]{\lam}}-\frac{3.26}{\lam^{3/4}}+\frac{2.53}{\lam}+
 \frac{4.03}{\lam^{5/4}}\\
 1.5 & 2
   \sqrt[4]{\lam}+\frac{2}{\sqrt[4]{\lam}}-\frac{3.27}{\lam^{3/4}}+\frac{2.63}{\lam}+
 \frac{3.83}{\lam^{5/4}}\\
 1.6 & 2
   \sqrt[4]{\lam}+\frac{2}{\sqrt[4]{\lam}}-\frac{3.28}{\lam^{3/4}}+\frac{2.70}{\lam}+
 \frac{3.69}{\lam^{5/4}}
\\\hline   
\end{array}~~~~
\eea
One can see that only $c_3$ is relatively stable $c_3=-3.23\pm0.05$ but the other coefficients change more substantially. 
\begin{figure}[t]
\begin{center}
\includegraphics*[width=0.45\textwidth]{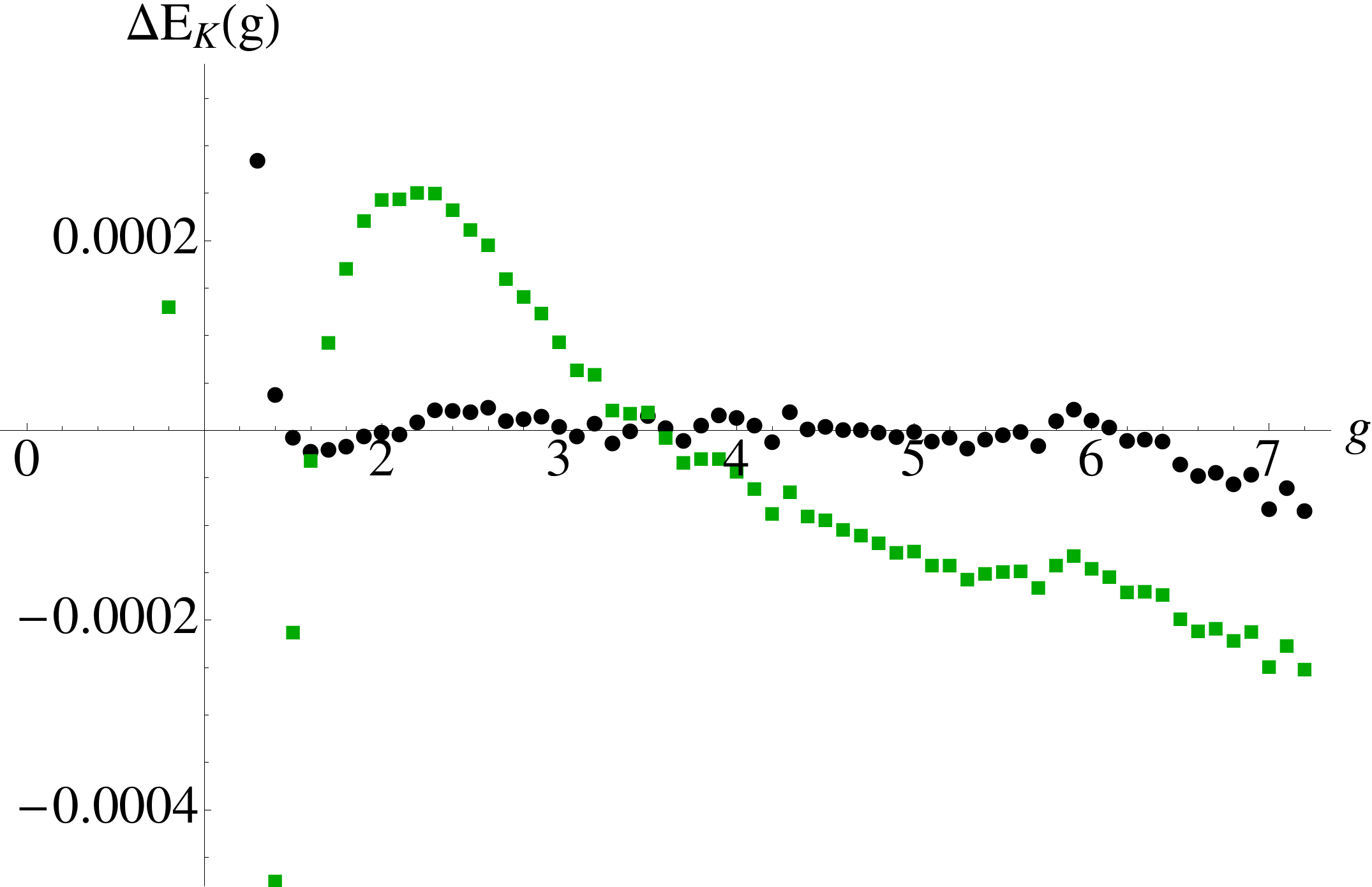}\quad \includegraphics*[width=0.45\textwidth]{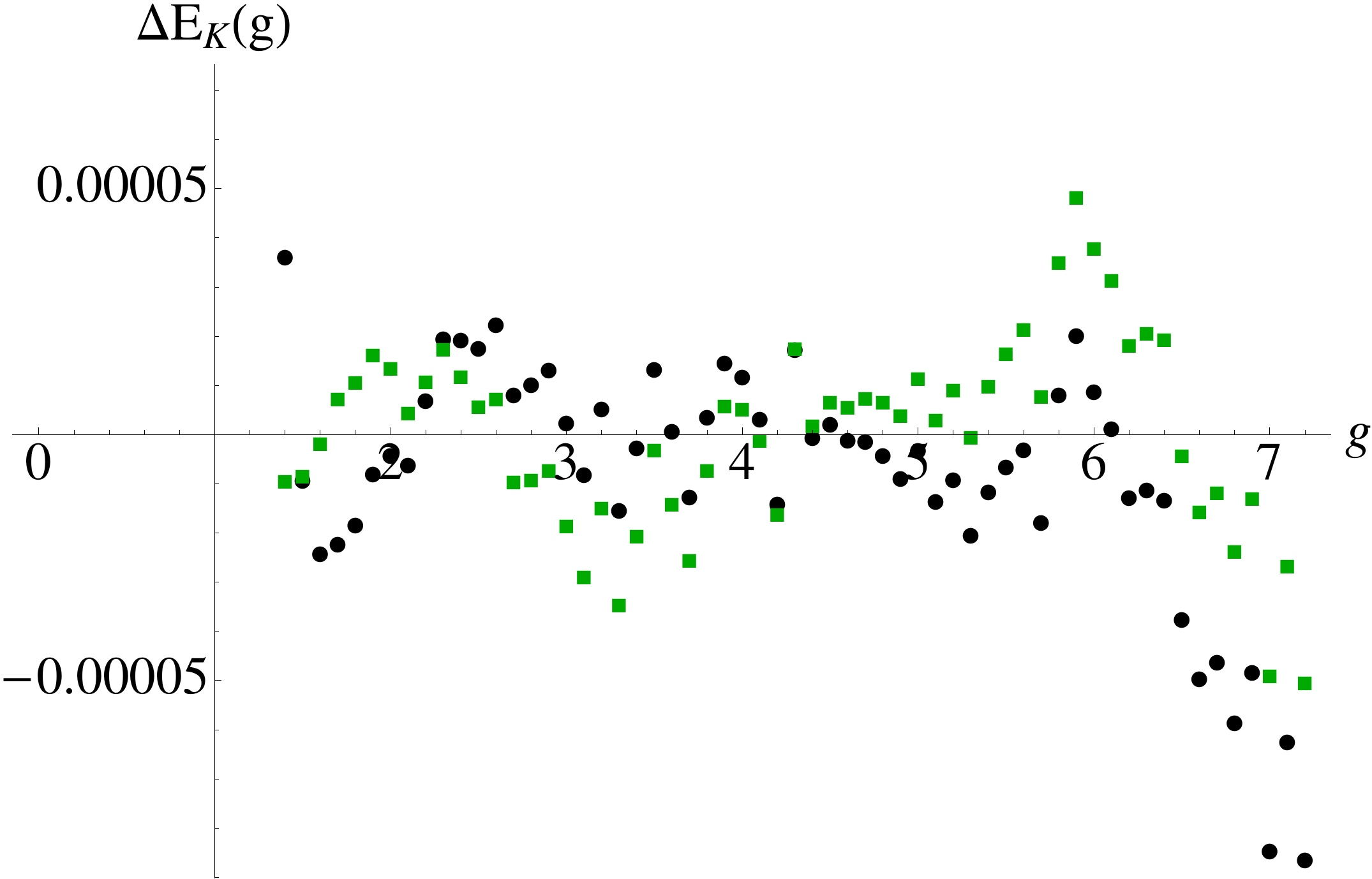}
\end{center}
\caption{\smaller Black dots represent the difference between the computed values of the Konishi energy and its fitting function (4.7): $\Delta E_K=E_K- {\overline E}_K$. Green squares on the left picture represent a similar difference with ${\overline E}_K^{\rm GKV}$ where the fitting function 
${\overline E}_K^{\rm GKV}(\lam)=2 \sqrt[4]{\lam}+\frac{2}{\sqrt[4]{\lam}}-\frac{2.94}{\lam^{3/4}}+\frac{8.83}{\lam^{5/4}}$ is obtained by using our data and setting $c_{-1}=c_1=2$ and $c_0=c_2=c_4=0$. On the right picture green squares represent the difference of $E_K$ and the fitting function in (1.2).}
\end{figure}  
\begin{figure}[t]
\begin{center}
\includegraphics*[width=0.47\textwidth]{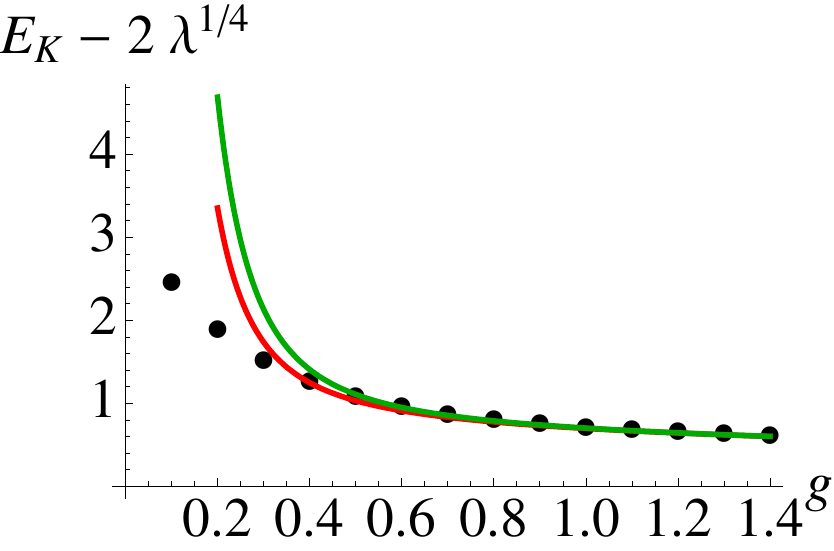}\quad \includegraphics*[width=0.47\textwidth]{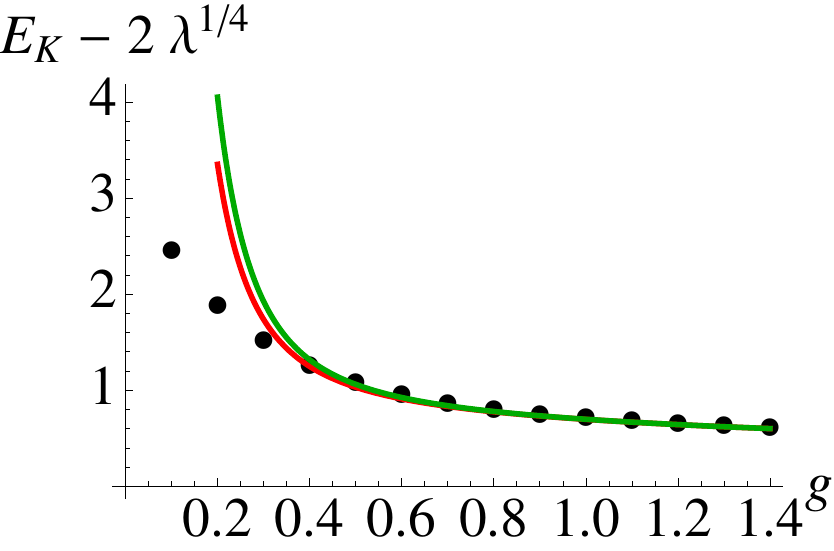}
\end{center}
\caption{\smaller  Black dots represent the difference between the energy and its leading large $\lambda$ asymptotic, $E_K- 2 \sqrt[4]{{\lam}}$. Red (upper) curve is the graph of ${\overline E}_K-2  \sqrt[4]{{\lam}}$. Green curve on the left picture is the graph of ${\overline E}_K^{\rm GKV}(\lam)-2  \sqrt[4]{{\lam}}$, and on the right picture it is the difference between  the fitting function (1.3) and $2 \sqrt[4]{{\lam}}$. The red curve works very well starting already with $g=0.3, \lam=3.55$.}
\end{figure}

Does our fitting rule out the strong coupling expansion in powers of $1/\sqrt\lam$ advocated in \cite{GKV09b}?
In Figure 23 we compare the fitting function ${\overline E}_K(\lam)$ \eqref{FFit14} with two fitting functions obtained by setting $c_{2k}=0$ from the very beginning. Both ${\overline E}_K(\lam)$ and the function \eqref{Emy} fit the data equally well. This is not surprising  because both functions have the same number of free fitting parameters.  On the other hand, the function ${\overline E}_K^{\rm GKV}(\lam)=2 \sqrt[4]{\lam}+\frac{2}{\sqrt[4]{\lam}}-\frac{2.94}{\lam^{3/4}}+\frac{8.83}{\lam^{5/4}}$ obtained by setting $c_{-1}=c_1=2$ and $c_{2k}=0$ significantly deviates from the data,  up to $0.0002$, which is at least twice more than the precision of our computation for $g<5$. Thus, we are to conclude that the coefficient $c_4$ does not vanish and  the strong coupling expansion is in powers of $1/\sqrt[4]\lam$. Having said this, we should admit that if 
the actual precision of our computation for $g\sim 7.0$ is not $0.0001$ but about $0.0002$, and if the contribution of exponentially suppressed terms is of order $0.0002$ for $g\sim 2$, then this could explain the left plot on Figure 23 and make possible the vanishing of $c_4$. 

\begin{figure}[t]
\begin{center}
\includegraphics*[width=0.6\textwidth]{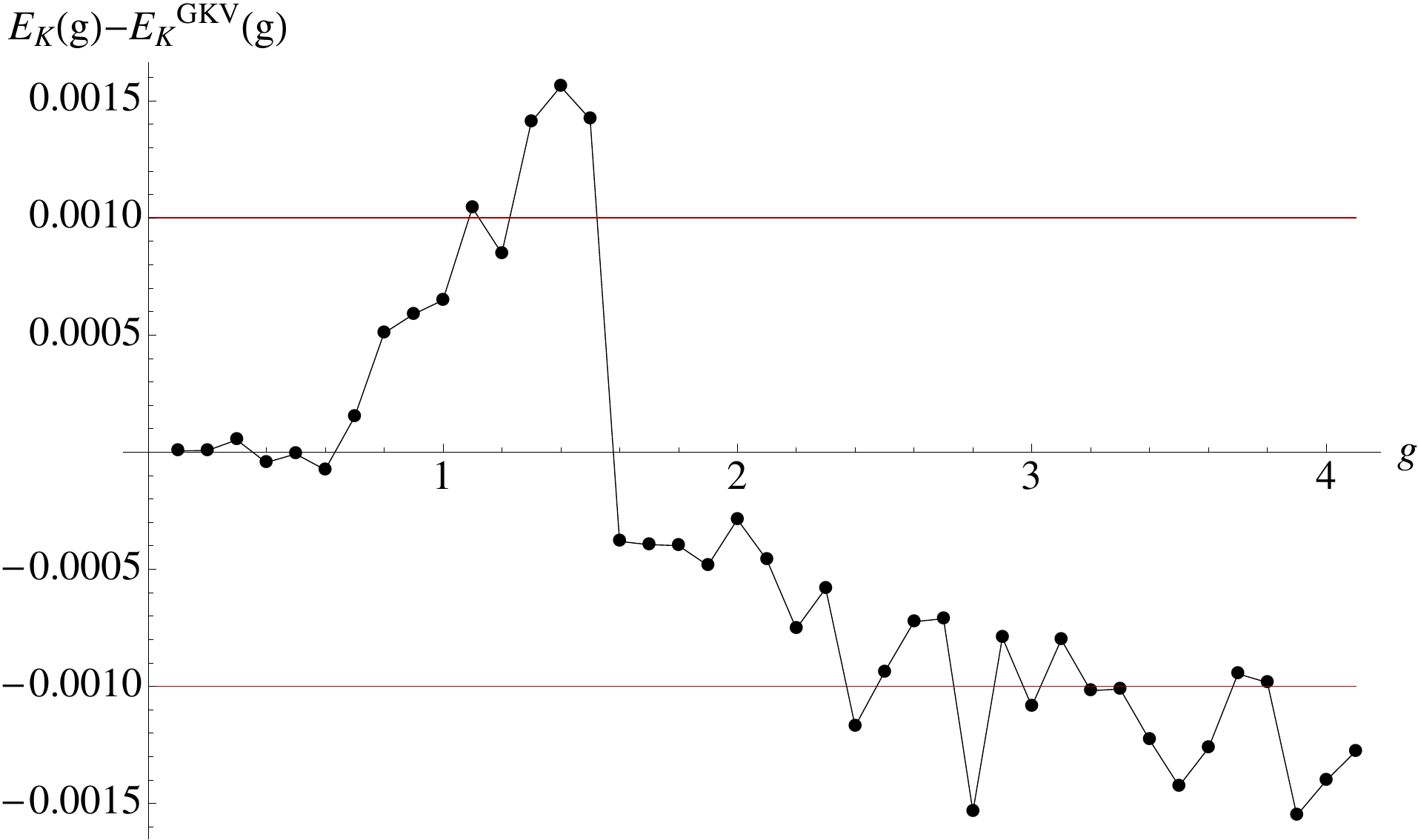}
\end{center}
\caption{\smaller The black dots represent the difference between the values of the Konishi energy we obtained and those from  \cite{GKV09b}.  One sees that our results agree with those of \cite{GKV09b} with $0.0015$ precision  for almost all values of $g$. 
}
\end{figure}

Let us finally mention that the function ${\overline E}_K$ works unexpectedly well
(and better than ${\overline E}_K^{\rm GKV}$ or \eqref{Emy}) starting already with such a small value of the coupling constant as 
$g=0.3, \lam=3.55$, see Figure 24, that is less than the expected radius of convergency of the weak-coupling expansion,  even though the fitting was done for the data with $g\ge 1.4$. This implicitly confirms that $c_1=2$.
Since $\sqrt[4]{3.55}=1.37$ is close to $1$, changing $c_1$ just by $0.1$ would require an essential change of the coefficients $c_3,c_4$ and $c_5$   
to fit the energy data at large $\lambda$ but then the fitting at small $\lambda$ would be destroyed.

In Figure 25 we compare our numerical results with those of \cite{GKV09b}.

\section{Conclusion}

In this paper we solved the TBA equations for the Konishi operator descendent  from the $\sl(2)$ sector proposed by Arutyunov, Suzuki and the author \cite{AFS09} up to 't Hooft's coupling $\lambda\approx 2046$. At this value 
the iterations converge very slowly.
 It would be important to improve the numerical algorithm and approach closer to the first critical value of $\lam$, and  go beyond it. One possible improvement would be to use Newton's method for solving the TBA equations as it was done for the Hubbard model in \cite{Takahashi}. Solving the TBA equations numerically for large values of $\lam$ is a challenging problem also because at large $\lam$ the Konishi state energy is very close to the energy of the other two Konishi-like states with $n=1$, see \cite{AFS09} for detail. It would be interesting to repeat the computation for these states, and for other Konishi-like states with larger string levels. 

Fitting the data for the Konishi state energy\footnote{We assumed the absence of $\log \lam$-dependent terms in the asymptotic large $\lam$ expansion. These terms are known to appear in the strong coupling expansion of  the Bethe-Yang equations \cite{Bec,RS09}.}  we found convincingly  that the first nontrivial subleading coefficient $c_1$ is equal to 2 and the coefficient $c_2$ vanishes in agreement with the previous prediction \cite{GKV09b}. The coefficient $c_4$ however does not vanish and therefore the strong coupling expansion seems to be not in powers of $1/\sqrt\lam$ (up to an overall $\sqrt[4]\lam$) but in powers of $1/\sqrt[4]\lam$.

 In principle increasing the values of $\lam$ may change the fitting coefficients and  invalidate the predictions for the coefficients. It is necessary to derive 
 them  by analytic means.

Extrapolating the results obtained shows  that the first critical value for the Konishi operator is $\lam_{cr}^{(1)} >5300$ and  it is probably in the range $5300<\lam_{cr}^{(1)} <5700$ which is significantly higher than the estimate based on the
large $J$ asymptotic solution that gives $\lam_{cr}^{(1)} \approx774$. 
This means that even at such a large value of $\lam$ we are still far from the strong coupling regime. This is an intermediate coupling regime and in fact it is the one where $Y_Q$-functions play the most important role. 
Moreover, if 
the contribution of $Y_Q$-functions to the Konishi state energy will continue growing as $\sqrt\lam$ then this would imply that at large $\lambda$ the exact Bethe root asymptotes to a constant less than 2, and  the critical values for the Konishi operator are absent. 

It is clear that the numerical computation we performed raised more questions than gave answers. Some of the questions can be answered only analytically, and we hope to address them in future.


\section*{Acknowledgements}
The author thanks Gleb Arutyunov, Zoltan Bajnok, Niklas Beisert, Davide Fioravanti, Tristan McLoughlin, Jan Plefka, Radu Roiban, Matthias Staudacher, Ryo Suzuki, Roberto Tateo, Arkady Tseytlin and Kostya Zarembo for interesting
discussions and comments on the manuscript, and Dmitri Grigoriev for computer help.  This work 
was supported in part by the Science Foundation Ireland under
Grants No. 07/RFP/PHYF104 and 09/RFP/PHY2142, and by  a
one-month Max-Planck-Institut f\"ur Gravitationsphysik
Albert-Einstein-Institut grant.

\section{Appendix}
\subsection*{Numerical data}
Here we collect our numerical data. 

In the table \eqref{Edata} we present the results of the computation of the energy of the Konishi state or, equivalently, the conformal dimension of the Konishi operator
as a function of $g$
{\smaller \bea\la{Edata}
\begin{array}{|c|c||c|c||c|c||c|c||c|c||c|c|}
\hline
g& {\rm E_K}&g&{\rm E_K} &g& {\rm E_K}&g&{\rm E_K}&g&{\rm E_K}&g&{\rm E_K} \\\hline
0.1 & 4.02971 &
 0.2 & 4.11551 &
 0.3 & 4.24885 &
 0.4 & 4.41886 &
 0.5 & 4.61469 &
 0.6 & 4.82682 \\
 0.7 & 5.04775 &
 0.8 & 5.27151 &
 0.9 & 5.49399 &
 1. & 5.71265 &
 1.1 & 5.92614 &
 1.2 & 6.13385 \\
 1.3 & 6.33561 &
 1.4 & 6.53156 &
 1.5 & 6.72212 &
 1.6 & 6.90752 &
 1.7 & 7.0881 &
 1.8 & 7.2642 \\
 1.9 & 7.43612 &
 2. & 7.60411 &
 2.1 & 7.76844 &
 2.2 & 7.92935 &
 2.3 & 8.08702 &
 2.4 & 8.24163 \\
 2.5 & 8.39336 &
 2.6 & 8.54238 &
 2.7 & 8.68879 &
 2.8 & 8.83276 &
 2.9 & 8.97441 &
 3. & 9.11381 \\
 3.1 & 9.2511 &
 3.2 & 9.38638 &
 3.3 & 9.51969 &
 3.4 & 9.65117 &
 3.5 & 9.78087 &
 3.6 & 9.90884 \\
 3.7 & 10.0352 &
 3.8 & 10.1599 &
 3.9 & 10.2831 &
 4. & 10.4049 &
 4.1 & 10.5252 &
 4.2 & 10.6442 \\
 4.3 & 10.7618 &
 4.4 & 10.8782 &
 4.5 & 10.9933 &
 4.6 & 11.1072 &
 4.7 & 11.22 &
 4.8 & 11.3316 \\
 4.9 & 11.4421 &
 5. & 11.5516 &
 5.1 & 11.66 &
 5.2 & 11.7675 &
 5.3 & 11.8739 &
 5.4 & 11.9794 \\
 5.5 & 12.084 &
 5.6 & 12.1877 &
 5.7 & 12.2905 &
 5.8 & 12.3924 &
 5.9 & 12.4936 &
 6. & 12.5938 \\
 6.1 & 12.6933 &
 6.2 & 12.792 &
 6.3 & 12.89 &
 6.4 & 12.9872 &
 6.5 & 13.0837 &
 6.6 & 13.1795 \\
 6.7 & 13.2746 &
 6.8 & 13.369 &
 6.9 & 13.4627 &
 7. & 13.5558 &
 7.1 & 13.6483 &
 7.2 & 13.7401
 \\\hline
\end{array}~~~
\eea
}
In the table \eqref{wdata}  we present the results of the computation of the Bethe root $w$
{\smaller \bea\la{wdata}
\begin{array}{|c|c||c|c||c|c||c|c||c|c||c|c|}
\hline
g& {w_K}&g&{w_K} &g& {w_K}&g&{w_K}&g&{ w_K}&g&{ w_K} \\\hline
0.1 & 5.88827 &
 0.2 & 3.11236 &
 0.3 & 2.25445 &
 0.4 & 1.87021 &
 0.5 & 1.67116 &
 0.6 & 1.56174 \\
 0.7 & 1.50066 &
 0.8 & 1.46819 &
 0.9 & 1.45317 &
 1. & 1.4489 &
 1.1 & 1.45121 &
 1.2 & 1.45758 \\
 1.3 & 1.46638 &
 1.4 & 1.47661 &
 1.5 & 1.48748 &
 1.6 & 1.49866 &
 1.7 & 1.50987 &
 1.8 & 1.52093 \\
 1.9 & 1.53174 &
 2. & 1.54224 &
 2.1 & 1.55239 &
 2.2 & 1.56217 &
 2.3 & 1.57159 &
 2.4 & 1.58064 \\
 2.5 & 1.58934 &
 2.6 & 1.59769 &
 2.7 & 1.60569 &
 2.8 & 1.61339 &
 2.9 & 1.62078 &
 3. & 1.62787 \\
 3.1 & 1.6347 &
 3.2 & 1.64127 &
 3.3 & 1.64758 &
 3.4 & 1.65367 &
 3.5 & 1.65953 &
 3.6 & 1.66517 \\
 3.7 & 1.67063 &
 3.8 & 1.67588 &
 3.9 & 1.68096 &
 4. & 1.68586 &
 4.1 & 1.6906 &
 4.2 & 1.69518 \\
 4.3 & 1.69963 &
 4.4 & 1.70392 &
 4.5 & 1.70809 &
 4.6 & 1.71212 &
 4.7 & 1.71603 &
 4.8 & 1.71983 \\
 4.9 & 1.72351 &
 5. & 1.72709 &
 5.1 & 1.73057 &
 5.2 & 1.73395 &
 5.3 & 1.73723 &
 5.4 & 1.74043 \\
 5.5 & 1.74354 &
 5.6 & 1.74657 &
 5.7 & 1.74952 &
 5.8 & 1.75239 &
 5.9 & 1.7552 &
 6. & 1.75793 \\
 6.1 & 1.76059 &
 6.2 & 1.76319 &
 6.3 & 1.76572 &
 6.4 & 1.7682 &
 6.5 & 1.77062 &
 6.6 & 1.77298 \\
 6.7 & 1.77529 &
 6.8 & 1.77755 &
 6.9 & 1.77975 &
 7. & 1.78191 &
 7.1 & 1.78402 &
 7.2 & 1.78608
 \\\hline
\end{array}~~~
\eea
}

\subsection*{Numerical algorithm }

We compute the Konishi state energy in several steps. At the first step one solves the TBA equations for a fixed Bethe root $w$ by iterations.  Equations for $Y_\pm$--functions are solved first, then equations for $Y_w$ and $Y_{vw}$, and finally equations for $Y_Q$.
For $g=0.7$ at the first iteration one uses the vacuum solution for $Y_w$-- and $Y_{vw}$--functions \cite{FS}, and asymptotic $Y_Q$-functions \cite{BJ08}.\footnote{For $g<0.7$ we just use the asymptotic solution for $Y_Q$-functions. It is sufficient for the precision we are after. For  small enough values of $g$ one can also use the asymptotic large $J$ solutions for Y-functions to start iterations. They are expressed in terms of transfer matrices  corresponding to various representations of the symmetry algebra of the model under consideration \cite{Kuniba,Tsuboi} (see also \cite{GKV09}). In the $\AdS$ case
the symmetry algebra of the light-cone string theory is the centrally extended $\su(2|2)$ superalgebra \cite{B05,AFPZ}, and explicit formulae for the transfer matrices
 were conjectured in \cite{B06} and derived in \cite{ALST}.   } 
For larger $g$ the solution found at the previous value of $g$, and the linear extrapolation of the Bethe root are used.

The number of iterations is bounded not to exceed $\overline N_{\rm iter}=10$, and the iterations also stop if the absolute value of the difference between the energies of two successive iterations, $dE_n=|E_n-E_{n-1}|$, becomes less than $de/10$ where $de=0.01$ is the precision the Konishi state energy is computed with at this step.

The number of $Y_Q$-functions computed is determined by the contribution of the asymptotic $Y_Q$-functions to the energy. It is given by 
$\int d\tilde p\, \log(1+Y_Q)$, and we solve the TBA equations only for those $Y_Q$-functions whose contributions exceed $10^{-5}$. 
We denote the total number of $Y_Q$-functions used in the computation by $Q_{\rm max}$, and it depends on the value of $g$, to be precise we used, $Q_{\rm max}=3$ for  $g=0.7$,  $Q_{\rm max}=4$ for  $0.8\le g\le 1.1$,  $Q_{\rm max}=5$ for  $1.2\le g\le 1.8$,  $Q_{\rm max}=6$ for  $1.9\le g\le  2.8$, 
$Q_{\rm max}=7$ for  $2.9\le g\le 4.2$, and 
$Q_{\rm max}=8$ for  $4.3\le g\le 7.2$. 

The number of $Y_w$-functions is equal to $N_{\rm w}=15$, and 
number of $Y_{vw}$-functions is equal to $N_{\rm w} + Q_{\rm max} -2$. This seems to be more than we need due to the locality of the simplified TBA equations for auxiliary functions, and because the hybrid  TBA equations for $Y_Q$-function involve only $Y_{1|vw}$- and $Y_{Q-1|vw}$-functions.

Y-functions are computed at discrete sets of points.
In particular, $Y_Q(u)$-functions are computed at 
points $u_k=k\, du\,,\ k=0,1,...$ with the step $du=10/1001$ in the interval  
$[0,u_{max}^{q}(Q)]$, where  $u_{max}^{q}(Q)$ is the lowest value of $u$ such that  $Y_Q^o(u_{max}^{q}(Q))<10^{-6}$. Then, one uses Mathematica function Interpolation, and also the large $u$ behavior of asymptotic $Y_Q$-functions to define $Y_Q$ as  continuous functions on the real $u$-line. 
$Y_w$-- (and $Y_{vw}$--functions) are computed until 
$|1-Y_{Q|w}(u_{max}^{w}(Q))/(Q(Q+2)) |<10^{-3}$ with a variable step equal to 
$du$ for $0 \le u<4$, to $2 du$ for $4 \le u< 12$ and so on. One uses again Interpolation and the large $u$ asymptotic of $Y_w$-- and $Y_{vw}$--functions
to define them for all $u$. Finally,
$Y_\pm(u)$-functions are computed at 
points $u_k=k\, du_y$ with the step $du_y=1/10$. 

At the second step  one uses the Y-functions to solve the exact Bethe equation, and find an adjusted value of $w$ and energy. Then one repeats the procedure until the difference between the energies becomes less than   $de=0.01$.

At the third step one uses the solution obtained to refine it by decreasing the steps $du$ and $du_y$, and the energy precision $de$. We first solved the TBA equations with $du\approx 1/50$, $du_y=1/50$ and $de =0.0001$.  Then we decreased the steps to $du\approx 1/100$, $du_y=1/100$. Finally we used the steps  $du\approx 1/130$, $du_y=1/130$ for $3.9\le g\le 4.9$, and 
$du\approx 1/150$, $du_y=1/150$ for $g\ge 5.0$.
The precision is about $0.0005$ for $g<5.0$. For larger values of $g$  the precision decreases and is about $0.001$. 

In version 2 of the paper to increase the precision we switched from the $u$ variable to the mirror momentum $\tilde p_Q$ rescaled so that the intervals $[0,2]$ are mapped to each other. We then used $g$-dependent steps $d\tilde p= 2/N_0 \sqrt{2/g}$ with $N_0=200$ and $N_0=250$. Finally, for $g>3$ we increased the number of auxiliary functions $N_{\rm w}=20$. 
Observing the changes in the energy, we believe that the final precision is about $0.0001$ for all values of $g$. 

To solve the hybrid TBA equations\footnote{To avoid computing many dressing kernels we first tried to use the simplified TBA equations for $Y_Q$-functions. It appeared however that they were numerically unstable (at least with our algorithm) and led to a large systematic error increasing with $g$. 
} 
we compute the dressing kernels $K^\Sigma_{Q1}(u,v)$ at the points $(k du, n du)\,,\ k,n=0,1,...$ with the step $du=50/1001$, and 
$K^\Sigma_{QQ'}(u,v)$, $Q'\ge 2$ with the step $du=100/1001$ in the rectangle $[0,u_{max}^{q}(Q)]\times [0,u_{max}^{q}(Q')]$, and $K^\Sigma_{Q1_*}(u,v)$ 
at the points $(k du,  \Delta v)\,,\ k=0,1,...$, where $\Delta v$ contains the Bethe root $w$, and the step in the $v$-direction is $dv=1/2000$.


\subsection*{Dressing phases and kernels}

The  improved dressing phases in various kinematic regions of the string and mirror models involve a common function which can be written as a sum of $\Phi$-functions, see \cite{AFdp} for definitions. 
We denote this function as
$\Theta(x_1^+\,,x_1^-\,,u_1\,,Q\,,x_2^+\,,x_2^-\,,u_2\,,M\,,g )$, 
where $x_i^\pm$ satisfy the usual  constraints
\bea\la{xpmQ}
x_1^\pm+{1\ov x_1^\pm} = u_1 \pm {i\ov g}Q\,,\quad  x_2^\pm+{1\ov x_2^\pm} = u_2 \pm {i\ov g}M\,.
\eea
If the parameters $x^\pm_i$ belong to the string region, that is $|x^\pm_i|>1$, then $\Theta$ is just equal to the BES dressing phase. 

One can show by using the DHM representation \cite{DHM} for the dressing phase that for any choice of $x^\pm_i$ satisfying  \eqref{xpmQ}, $\Theta$ has an integral representation involving elliptic $\Pi$-function. Since the derivation of the representation is rather involved we present below the expression leaving a proof of the formula to curious postdocs
{\smaller
\bea\la{drph}
&&\Theta(x_1^+\,,x_1^-\,,u_1\,,Q\,,x_2^+\,,x_2^-\,,u_2\,,M\,,g )=\frac{i g^2}{\pi ^2 } \int_{0}^1\, 
\frac{dt}{2-t} \log{\Gamma(1-2 i g (1-t))\ov \Gamma(1+2 i g (1-t))}\Big[
\\\nonumber
&&-\frac{i \left((x_1^-)^2-1\right) (x_2^--x_2^+) (x_2^- x_2^+ Q+Q+i g (x_2^-
   x_2^+ u_1+u_1-2 (2 x_2^- x_2^+ (1-t)+2 (1-t)+x_2^-+x_2^+)))}{(-i Q-2 g (1-t)+g u_1) \left(M^2-(Q+i g (-4 (1-t)+u_1-u_2))^2\right) x_1^- x_2^- x_2^+}  
  \\ \nonumber
  &&  \times \Pi \Big(\frac{4 g^2 t^2}{(-i Q-2 g (1-t)+g u_1)^2}|\frac{t^2}{(2-t)^2}\Big)
  \\\nonumber
  &&
  -\frac{i \left((x_1^+)^2-1\right) (x_2^--x_2^+) (x_2^- x_2^+ Q+Q-i g (x_2^- x_2^+ u_1+u_1-2 (2 x_2^- x_2^+ (1-t)+2 (1-t)+x_2^-+x_2^+))) }{(i Q-2 g (1-t)+g u_1) \left(M^2-(Q-i g (-4
   (1-t)+u_1-u_2))^2\right) x_2^- x_1^+ x_2^+}
  \\ \nonumber
  && 
   \times  \Pi \Big(\frac{4 g^2 t^2}{(i Q-2 g (1-t)+g
   u_1)^2}|\frac{t^2}{(2-t)^2}\Big)
   \\\nonumber
   &&
    -\frac{\left((x_1^+)^2-1\right)
   (x_2^--x_2^+) (i Q (x_2^- x_2^++1)+g (4 (1-t)+u_1-2 x_2^-+(4 (1-t) x_2^-+u_1
   x_2^--2) x_2^+)) }{(-i Q-g (2
   (1-t)+u_1)) \left(M^2-(Q-i g (4 (1-t)+u_1-u_2))^2\right) x_2^- x_1^+
   x_2^+}
  \\ \nonumber
  && 
   \times  \Pi \Big(\frac{4 g^2 t^2}{(-i Q-g (2 (1-t)+u_1))^2}|\frac{t^2}{(2-t)^2}\Big)
     \\\nonumber
   &&
   +\frac{\left((x_1^-)^2-1\right) (x_2^--x_2^+) (g (-4 t+u_1-2 x_2^-+(4 (1-t)
   x_2^-+u_1 x_2^--2) x_2^++4)-i Q (x_2^- x_2^++1)) }{(i Q-g (2 (1-t)+u_1)) \left(M^2-(Q+i g (4
   (1-t)+u_1-u_2))^2\right) x_1^- x_2^- x_2^+}
  \\ \nonumber
  && 
   \times  \Pi \Big(\frac{4 g^2 t^2}{(i Q-g (2
   (1-t)+u_1))^2}|\frac{t^2}{(2-t)^2}\Big)
  \\ \nonumber
  && 
   -\frac{\left((x_2^-)^2-1\right)
   (x_1^--x_1^+) (g (x_1^- x_1^+ u_2+u_2-2 (2 x_1^- x_1^+ (1-t)+2
   (1-t)+x_1^-+x_1^+))-i M (x_1^- x_1^++1)) }{\left(Q^2-(M-i g (4 (1-t)+u_1-u_2))^2\right) (-i M-2 g (1-t)+g
   u_2) x_1^- x_2^- x_1^+}
  \\ \nonumber
  && 
    \times \Pi \Big(\frac{4 g^2 t^2}{(-i M-2 g (1-t)+g
   u_2)^2}|\frac{t^2}{(2-t)^2}\Big)
  \\ \nonumber
  && 
   +\frac{(x_1^--x_1^+) (i M (x_1^- x_1^++1)+g (x_1^-
   x_1^+ u_2+u_2-2 (2 x_1^- x_1^+ (1-t)+2 (1-t)+x_1^-+x_1^+))) \left((x_2^+)^2-1\right)}{\left(Q^2-(M+i g (4
   (1-t)+u_1-u_2))^2\right) (i M-2 g (1-t)+g u_2) x_1^- x_1^+
   x_2^+}
  \\ \nonumber
  && 
    \times  \Pi \Big(\frac{4 g^2 t^2}{(i M-2 g (1-t)+g u_2)^2}|\frac{t^2}{(2-t)^2}\Big)
   \\ \nonumber
  &&  
   +\frac{\left((x_2^-)^2-1\right) (x_1^--x_1^+) (g (4 (1-t)+u_2-2 x_1^-+(4 (1-t)
   x_1^-+u_2 x_1^--2) x_1^+)-i M (x_1^- x_1^++1))}{\left(Q^2-(M-i g (-4 (1-t)+u_1-u_2))^2\right) (-i M+2 g (1-t)+g
   u_2) x_1^- x_2^- x_1^+}
  \\ \nonumber
  && 
    \times  \Pi \Big(\frac{4 g^2 t^2}{(-i M+2 g (1-t)+g
   u_2)^2}|\frac{t^2}{(2-t)^2}\Big)
  \\ \nonumber
  &&  
   -\frac{(x_1^--x_1^+) (i M (x_1^- x_1^++1)+g (4
   (1-t)+u_2-2 x_1^-+(4 (1-t) x_1^-+u_2 x_1^--2) x_1^+)) \left((x_2^+)^2-1\right) }{\left(Q^2-(M+i g (-4
   (1-t)+u_1-u_2))^2\right) (i M+2 g (1-t)+g u_2) x_1^- x_1^+ x_2^+}
  \\ \nonumber
  && 
   \times  \Pi
   \Big(\frac{4 g^2 t^2}{(i M+2 g (1-t)+g u_2)^2}|\frac{t^2}{(2-t)^2}\Big) 
   \Big] \,.
\eea
}
This integral representation significantly speeds up computing the dressing phases.
For small values of $g$ it might be useful to do the change $t\to (1-t)/g$ but for very large $g$ the representation seems to be  the best one for numerics. 

The function $\Theta$ is then used to define the corresponding expressions 
$\Theta_{ss}$, $\Theta_{sm}$, $\Theta_{ms}$ and $\Theta_{mm}$  in different regions of the $z$-torus
{\smaller
\bea\la{thss}
&&\Theta_{ss}(u,Q,v,M,g)= \Theta \left(x_s\big(u+\frac{i
   Q}{g}\big),x_s\big(u-\frac{i
   Q}{g}\big),u,Q,x_s\big(v+\frac{i
   M}{g}\big),x_s\big(v-\frac{i M}{g}\big),v,M,g\right)~~~~~\\
   \nonumber
&&\Theta_{sm}(u,Q,v,M,g)= \Theta \left(x_s\big(u+\frac{i
   Q}{g}\big),x_s\big(u-\frac{i
   Q}{g}\big),u,Q,x\big(v+\frac{i
   M}{g}\big),x\big(v-\frac{i M}{g}\big),v,M,g\right)\\
   \nonumber
&&\Theta_{ms}(u,Q,v,M,g)= \Theta \left(x\big(u+\frac{i
   Q}{g}\big),x\big(u-\frac{i
   Q}{g}\big),u,Q,x_s\big(v+\frac{i
   M}{g}\big),x_s\big(v-\frac{i M}{g}\big),v,M,g\right)\\
   \nonumber
&&\Theta_{mm}(u,Q,v,M,g)= \Theta \left(x\big(u+\frac{i
   Q}{g}\big),x\big(u-\frac{i
   Q}{g}\big),u,Q,x\big(v+\frac{i
   M}{g}\big),x\big(v-\frac{i M}{g}\big),v,M,g\right)
\eea
}
Here $x(u)$ and $x_s(u)$ are the mirror and string $x$-functions:
$x(u) = {1\ov 2}(u - i\sqrt{4-u^2})$, and $x_s(u) = {1\ov 2}u(1 +\sqrt{1-4/u^2})$.
In particular, $\Theta_{ss}$ is just the BES dressing phase.

The improved dressing phases also involve contributions of $\Psi$-functions \cite{AFdp}. 
The total contribution of $\Psi(x_2,x_1)$-functions to the improved dressing phase in the string-mirror region  is given by
{\smaller
\bea\la{psi21}
&&\Psi_{21}(x_1^+,x_1^-,u_2,g,M)= \int_0^{\pi/2} d\theta\, \sin (\theta ) \Big[ \\\nonumber
&&\frac{i g M}{\pi  \left(4 \left(\frac{g u_2}{2}-g \cos
   (\theta )\right)^2+M^2\right)} \log \left(\frac{-i (x_1^--x_1^+)
   \sin (\theta )-(x_1^-+x_1^+) \cos (\theta )+x_1^-
   x_1^++1}{i (x_1^--x_1^+) \sin (\theta)
  -(x_1^-+x_1^+) \cos (\theta )+x_1^-
   x_1^++1}\right)
   \\\nonumber
   &&
   -\frac{2 M}{4 \left(\frac{g
   u_2}{2}+g \cos (\theta )\right)^2+M^2} \log
   \left(\frac{-i (x_1^--x_1^+) \sin (\theta
   )+(x_1^-+x_1^+) \cos (\theta )+x_1^- x_1^++1}{i
   (x_1^--x_1^+) \sin (\theta )+(x_1^-+x_1^+) \cos
   (\theta )+x_1^- x_1^++1}\right)
   \\\nonumber
   &&
   + \left(\psi
   ^{(0)}\left(\frac{M}{2}+\frac{i g u_2}{2}-i g \cos (\theta
   )+1\right)+\psi ^{(0)}\left(\frac{M}{2}-\frac{i g u_2}{2}+i g
   \cos (\theta )+1\right)\right) 
   \\\nonumber
   &&
   ~~\times \log \left(\frac{-i
   (x_1^--x_1^+) \sin (\theta )-(x_1^-+x_1^+) \cos
   (\theta )+x_1^- x_1^++1}{i (x_1^--x_1^+) \sin
   (\theta )-(x_1^-+x_1^+) \cos (\theta )+x_1^-
   x_1^++1}\right)
   \\\nonumber
   &&
   + \left(\psi
   ^{(0)}\left(\frac{M}{2}-\frac{i g u_2}{2}-i g \cos (\theta
   )+1\right)+\psi ^{(0)}\left(\frac{M}{2}+\frac{i g u_2}{2}+i g
   \cos (\theta )+1\right)\right) 
    \\\nonumber
   &&
  ~~ \times \log \left(\frac{-i
   (x_1^--x_1^+) \sin (\theta )+(x_1^-+x_1^+) \cos
   (\theta )+x_1^- x_1^++1}{i (x_1^--x_1^+) \sin
   (\theta )+(x_1^-+x_1^+) \cos (\theta )+x_1^-
   x_1^++1}\right)\Big]
\eea
}

Here the definitions  are the same as above but $Q=1$, and $x_1^\pm$ are in the string region, that is $x_1^\pm=x_s(u_1\pm{i\ov g})$, while $x_2^\pm$ are in the mirror region, that is $x_2^\pm=x(u_2\pm{i\ov g}M)$. It is convenient to introduce
{\smaller
\bea\la{psi21b}
\Psi_{21}^{\rm sm}(u_1,Q,u_2,M,g)=\Psi_{21}\left(x_s\big(u_1+\frac{i Q}{g}\big),x_s\big(u_1-\frac{i Q}{g}\big),u_2,g,M\right)
\eea
}
Then, the contribution of the last term in the formula for the  improved string-mirror dressing phase is given by
{\smaller
\bea\la{lastsm}
&&\frac{1}{i}\log\Sigma_{\rm last}^{\rm sm}(u_1,Q,u_2,M,g) =\\\nonumber
&&\frac{1}{2i}  \log \frac{\left(x_s\left(u_1-\frac{i
   Q}{g}\right)-\frac{1}{x\left(u_2-\frac{i M}{g}\right)}\right)
   \left(x_s\left(u_1-\frac{i
   Q}{g}\right)-x\left(u_2+\frac{i M}{g}\right)\right)
   \left(x_s\left(u_1+\frac{i
   Q}{g}\right)-\frac{1}{x\left(u_2-\frac{i
   M}{g}\right)}\right)}{\left(x_s\left(u_1-\frac{i
   Q}{g}\right)-\frac{1}{x\left(u_2+\frac{i M}{g}\right)}\right)^2
   \left(x_s\left(u_1+\frac{i
   Q}{g}\right)-x\left(u_2+\frac{i M}{g}\right)\right)}\
\eea
}
The total improved dressing phase ${1\ov i}\log \Sigma_{1_*M}$ in the string-mirror region is thus given by the sum of the three terms
{\smaller
\bea\nonumber
{1\ov i}\log \Sigma_{1_*M}(u_1,1,u_2,M,g) &=& \Theta^{\rm sm}(u_1,1,u_2,M,g) +\Psi_{21}^{\rm sm}(u_1,1,u_2,M,g)\\\la{drsm}
&+&\frac{1}{i}\log\Sigma_{\rm last}^{\rm sm}(u_1,1,u_2,M,g) 
\eea
}

The improved dressing phase in the mirror-string region is related to the one in the string-mirror region by the unitarity relation
\bea
{1\ov i}\log \Sigma_{M1_*}(u_2,u_1)=-{1\ov i}\log \Sigma_{1_*M}(u_1,u_2)\,.
\eea

\bigskip

Finally, the improved dressing phase in the mirror-mirror region is given by the sum of four terms. The first one is $\Theta_{mm}$. 
The second term is  the total contribution of $\Psi(x_2,x_1)$-functions  given by
{\smaller
\bea\la{psi21mm}
&&\Psi_{21}^{\rm m}(x_1^+,x_1^-,u_2,g,M) = \frac{i}{2} 
\log{\Gamma\left(-i g+\frac{1}{2} i (gu_2-i M)+1\right)\Gamma\left(-i g+\frac{1}{2} i (i
   M+g u_2)+1\right)\ov\Gamma \left(i g-\frac{1}{2} i (g
   u_2-i M)+1\right)\Gamma\left(i g-\frac{1}{2} i (i
   M+g u_2)+1\right)}
   \\\nonumber
   &&
~~~~~+
 \frac{g} {2 \pi }\int_0^{\pi/2} d\theta\, \sin (\theta )\left(-i \log \left(\frac{1-\frac{e^{-i \theta
   }}{x_1^-}}{1-\frac{e^{i \theta }}{x_1^-}}\right)-i \log
   \left(\frac{1-x_1^+ e^{-i \theta }}{1-x_1^+ e^{i \theta
   }}\right)+2 \theta -2 \pi \right)\\\nonumber
   &&
 \times   \left(-\frac{2 M}{4 \left(\frac{g
   u_2}{2}-g \cos (\theta )\right)^2+M^2}+\psi
   ^{(0)}\left(\frac{M}{2}+\frac{i g u_2}{2}-i g \cos (\theta
   )+1\right)+\psi ^{(0)}\left(\frac{M}{2}-\frac{i g u_2}{2}+i g
   \cos (\theta )+1\right)\right)
   \\\nonumber
   &&
 ~~~~  +\sin (\theta ) \left(-i \log
   \left(\frac{1+\frac{e^{i \theta }}{x_1^-}}{1+\frac{e^{-i \theta
   }}{x_1^-}}\right)-i \log \left(\frac{1+x_1^+ e^{i \theta
   }}{1+x_1^+ e^{-i \theta }}\right)-2 \theta \right)
   \\\nonumber
   &&
    \times   
    \left(-\frac{2
   M}{4 \left(\frac{g u_2}{2}+g \cos (\theta )\right)^2+M^2}+\psi
   ^{(0)}\left(\frac{M}{2}-\frac{i g u_2}{2}-i g \cos (\theta
   )+1\right)+\psi ^{(0)}\left(\frac{M}{2}+\frac{i g u_2}{2}+i g
   \cos (\theta )+1\right)\right)
\eea
}

The third term is  the total contribution of $\Psi(x_1,x_2)$-functions   given by
{\smaller
\bea\la{psi12mm} 
&&\Psi_{12}^{\rm m}(u_1,x_2^+,x_2^-,g,M) = 
-\frac{i}{2} \log{\Gamma\left(-i g+\frac{1}{2} i (g
   u_1-i M)+1\right)\Gamma\left(-i g+\frac{1}{2} i (i
   M+g u_1)+1\right)\ov\Gamma\left(i g-\frac{1}{2} i (g
   u_1-i M)+1\right)\Gamma\left(i g-\frac{1}{2} i (i
   M+g u_1)+1\right)}\\\nonumber
   &&
  ~~~~ + \frac{g} {2 \pi }\int_0^{\pi/2} d\theta\, \sin (\theta ) \left(i \log \left(\frac{1-\frac{e^{-i
   \theta }}{x_2^-}}{1-\frac{e^{i \theta }}{x_2^-}}\right)+i
   \log \left(\frac{1-x_2^+ e^{-i \theta }}{1-x_2^+ e^{i \theta
   }}\right)-2 \theta +2 \pi \right) \\\nonumber
   &&\times 
   \left(-\frac{2 M}{4 \left(\frac{g
   u_1}{2}-g \cos (\theta )\right)^2+M^2}+\psi
   ^{(0)}\left(\frac{M}{2}+\frac{i g u_1}{2}-i g \cos (\theta
   )+1\right)+\psi ^{(0)}\left(\frac{M}{2}-\frac{i g u_1}{2}+i g
   \cos (\theta )+1\right)\right)
   \\\nonumber
   &&
   ~~~~+\sin (\theta ) \left(i \log
   \left(\frac{1+\frac{e^{i \theta }}{x_2^-}}{1+\frac{e^{-i \theta
   }}{x_2^-}}\right)+i \log \left(\frac{1+x_2^+ e^{i \theta
   }}{1+x_2^+ e^{-i \theta }}\right)+2 \theta \right)
    \\\nonumber
   &&\times
    \left(-\frac{2
   M}{4 \left(\frac{g u_1}{2}+g \cos (\theta )\right)^2+M^2}+\psi
   ^{(0)}\left(\frac{M}{2}-\frac{i g u_1}{2}-i g \cos (\theta
   )+1\right)+\psi ^{(0)}\left(\frac{M}{2}+\frac{i g u_1}{2}+i g
   \cos (\theta )+1\right)\right)
\eea
}

We also introduce
{\smaller
\bea\la{psimms}
&&\Psi_{12}^{\rm mm}(u_1, Q, u_2, M, g)= \Psi_{12}^{\rm m}\left(u_1,x\left(u_2+\frac{i
   M}{g}\right),x\left(u_2-\frac{i M}{g}\right),g,Q\right)\\\nonumber
   &&
   \Psi_{21}^{\rm mm}(u_1, Q, u_2, M, g)=  \Psi_{21}^{\rm m}\left(x\left(u_1+\frac{i
   Q}{g}\right),x\left(u_1-\frac{i Q}{g}\right),u_2,g,M\right)
\eea
}

Finally, the last term is given by 
{\smaller
\bea\la{lastmm}
&&\frac{1}{i}\log\Sigma_{\rm last}^{\rm mm}(u_1,Q,u_2,M,g) =\\\nonumber
&&{1\ov i} \log \left(\frac{i^{Q}\Gamma
   \left(M-\frac{1}{2} i (i (Q-M)+g u_1-g
   u_2)\right) \left(1-\frac{1}{x\left(u_2-\frac{i
   M}{g}\right) x\left(u_1+\frac{i Q}{g}\right)}\right)
  }{i^{M}\Gamma
   \left(Q+\frac{1}{2} i (i (Q-M)+g u_1-g u_2)\right)\left(1-\frac{1}{x\left(u_2+\frac{i
   M}{g}\right) x\left(u_1-\frac{i Q}{g}\right)}\right) } \sqrt{\frac{x\left(u_2-\frac{i M}{g}\right)
   x\left(u_1+\frac{i Q}{g}\right)}{x\left(u_2+\frac{i
   M}{g}\right) x\left(u_1-\frac{i Q}{g}\right)}} \right)
\eea
}

Thus, the improved dressing phase $\Sigma_{QM}$ in the mirror-mirror region is given by
{\smaller
\bea\nonumber
{1\ov i}\log \Sigma_{QM}(u_1,Q,u_2,M,g) &=& \Theta^{\rm mm}(u_1,Q,u_2,M,g) +\Psi_{12}^{\rm mm}(u_1,Q,u_2,M,g)+\Psi_{21}^{\rm mm}(u_1,Q,u_2,M,g)\\\la{phmm}
&+&\frac{1}{i}\log\Sigma_{\rm last}^{\rm mm}(u_1,Q,u_2,M,g) 
\eea
}

The formulae above are valid for any real $u_1,u_2$, and should be used with caution for complex values of the parameters.

Finally, the dressing kernels are defined as 
\bea\nonumber
K^\Sigma_{QM}(u,v)={1\ov 2\pi i}{d\ov du}\log \Sigma_{QM}(u,v)\,,\quad K^\Sigma_{Q1_*}(u,v)={1\ov 2\pi i}{d\ov du}\log\Sigma_{Q1_*}(u,v)\,.
\eea

\subsection*{Mathematica form of the dressing phases}

To make sure there are no misprints in the formulae above we 
 also present below the same expressions in Mathematica input form.
 
 Eq. \eqref{drph} takes the form
{\smaller
\begin{verbatim}
\[CapitalTheta][xp1_, xm1_, u1_, Q_, xp2_, xm2_, u2_, M_, g_] :=
(1/Pi^2)*I*g^2*NIntegrate[((((-1+xm1^2)*(xm2-xp2))/(xm1*xm2*xp2))*
(((-I)*Q*(1+xm2*xp2)+g*(4-4*t+u1-2*xm2+(-2+4*(1-t)*xm2+u1*xm2)*xp2))/
((I*Q-g*(2*(1-t)+u1))*(M^2-(Q+I*g*(4*(1-t)+u1-u2))^2)))*
EllipticPi[((2*g*t)/(I*Q-g*(2*(1-t)+u1)))^2,(t/(2-t))^2]+
(-(((-1+xp1^2)*(xm2-xp2))/(xm2*xp1*xp2)))*((I*Q*(1+xm2*xp2)+
g*(4*(1-t)+u1-2*xm2+(-2+4*(1-t)*xm2+u1*xm2)*xp2))/(((-I)*Q-g*(2*(1-t)+u1))*
(M^2-(Q-I*g*(4*(1-t)+u1-u2))^2)))*
EllipticPi[((2*g*t)/((-I)*Q-g*(2*(1-t)+u1)))^2,(t/(2-t))^2]
+(-((I*(-1+xp1^2)*(xm2-xp2))/(xm2*xp1*xp2)))*
((Q+Q*xm2*xp2-I*g*(u1+u1*xm2*xp2-2*(2*(1-t)+xm2+xp2+2*(1-t)*xm2*xp2)))/
((I*Q-2*g*(1-t)+g*u1)*(M^2-(Q-I*g*(-4*(1-t)+u1-u2))^2)))*
EllipticPi[((2*g*t)/(I*Q-2*g*(1-t)+g*u1))^2,(t/(2-t))^2]+
(-((I*(-1+xm1^2)*(xm2-xp2))/(xm1*xm2*xp2)))*
((Q+Q*xm2*xp2+I*g*(u1+u1*xm2*xp2-2*(2*(1-t)+xm2+xp2+2*(1-t)*xm2*xp2)))/
(((-I)*Q-2*g*(1-t)+g*u1)*(M^2-(Q+I*g*(-4*(1-t)+u1-u2))^2)))*
EllipticPi[((2*g*t)/((-I)*Q-2*g*(1-t)+g*u1))^2,(t/(2-t))^2]+
(-(((xm1-xp1)*(-1+xp2^2))/(xm1*xp1*xp2)))*((I*M*(1+xm1*xp1)+
g*(4*(1-t)+u2-2*xm1+(-2+4*(1-t)*xm1+u2*xm1)*xp1))/((I*M+2*g*(1-t)+g*u2)*
(Q^2-(M+I*g*(-4*(1-t)+u1-u2))^2)))*
EllipticPi[((2*g*t)/(I*M+2*g*(1-t)+g*u2))^2,(t/(2-t))^2]
+(((-1+xm2^2)*(xm1-xp1))/(xm1*xm2*xp1))*
(((-I)*M*(1+xm1*xp1)+g*(4*(1-t)+u2-2*xm1+(-2+4*(1-t)*xm1+u2*xm1)*xp1))/
((Q^2-(M-I*g*(-4*(1-t)+u1-u2))^2)*((-I)*M+2*g*(1-t)+g*u2)))*
EllipticPi[((2*g*t)/((-I)*M+2*g*(1-t)+g*u2))^2,(t/(2-t))^2]+
(((xm1-xp1)*(-1+xp2^2))/(xm1*xp1*xp2))*((I*M*(1+xm1*xp1)+
g*(u2+u2*xm1*xp1-2*(2*(1-t)+xm1+xp1+2*(1-t)*xm1*xp1)))/((I*M-2*g*(1-t)+g*u2)*
(Q^2-(M+I*g*(4*(1-t)+u1-u2))^2)))*
EllipticPi[((2*g*t)/(I*M-2*g*(1-t)+g*u2))^2,(t/(2-t))^2]
+(-(((-1+xm2^2)*(xm1-xp1))/(xm1*xm2*xp1)))*
(((-I)*M*(1+xm1*xp1)+g*(u2+u2*xm1*xp1-2*(2*(1-t)+xm1+xp1+2*(1-t)*xm1*xp1)))/
(((-I)*M-2*g*(1-t)+g*u2)*(Q^2-(M-I*g*(4*(1-t)+u1-u2))^2)))*
EllipticPi[((2*g*t)/((-I)*M-2*g*(1-t)+g*u2))^2,(t/(2-t))^2])*
((LogGamma[1-2*I*g*(1-t)]-LogGamma[1+2*I*g*(1-t)])/(2-t)),{t,0,1}];
\end{verbatim}
}
Here xp1$\ \equiv x_1^+$, and so on, $x_1^++1/x_1^+=u_1+{i\ov g}Q$, and 
$x_2^++1/x_2^+=u_2+{i\ov g}M$.
 For effective numerics one should exclude the point $t=1 - {1\ov 4}|u_1 - u_2|$ (if $|u_1 - u_2| < 4$) from the integration region. This doubles the formula above, and by this reason we leave the necessary modification of the formula to the interested reader. For some points close to $\pm 2$ the formula works only if one specifies WorkingPrecision. For small values of $g$ and for the precision we are after it is sufficient to use    WorkingPrecision$\to$10.

$\Theta_{ss}$, $\Theta_{sm}$, $\Theta_{ms}$ and $\Theta_{mm}$ in \eqref{thss}
are written as
{\smaller
\begin{verbatim}
\[CapitalTheta]ss[u_, Q_, v_, M_, g_] := 
\[CapitalTheta][xs[u+(I/g)Q],xs[u-(I/g)Q],u,Q,xs[v+(I/g)M],xs[v-(I/g)M],v,M,g]; 
\[CapitalTheta]sm[u_, Q_, v_, M_, g_] := 
\[CapitalTheta][xs[u+(I/g)Q],xs[u-(I/g)Q],u,Q,x[v+(I/g)M],x[v-(I/g)M],v,M,g]; 
\[CapitalTheta]ms[u_, Q_, v_, M_, g_] := 
\[CapitalTheta][x[u+(I/g)Q],x[u-(I/g)Q],u,Q,xs[v+(I/g)M],xs[v-(I/g)M],v,M,g]; 
\[CapitalTheta]mm[u_, Q_, v_, M_, g_] := 
\[CapitalTheta][x[u+(I/g)Q],x[u-(I/g)Q],u,Q,x[v+(I/g)M],x[v-(I/g)M],v,M,g];
 \end{verbatim}
}
Here $x[u]$ and $xs[u]$ are the mirror and string $x$-functions.

The total contribution of $\Psi(x_2,x_1)$-functions to the improved dressing phase in the string-mirror region  in \eqref{psi21} is given by
{\smaller
\begin{verbatim}
\[CapitalPsi]21[xp1_,xm1_,u2_,g_,M_]:= g/(2 \[Pi] I) (NIntegrate[ 
Sin[\[Theta]](Log[(1+xm1 xp1-(xm1+xp1) Cos[\[Theta]]
-I (xm1-xp1) Sin[\[Theta]])/(1+xm1 xp1-(xm1+xp1) 
Cos[\[Theta]]+I (xm1-xp1)Sin[\[Theta]])]((-2 M)/(M^2+4((g u2)/2-g Cos[\[Theta]])^2)))
+ Sin[\[Theta]](Log[(1+xm1 xp1+(xm1+xp1) Cos[\[Theta]]
-I (xm1-xp1) Sin[\[Theta]])/(1+xm1 xp1+(xm1+xp1) 
Cos[\[Theta]]+I (xm1-xp1)Sin[\[Theta]])]((-2 M)/(M^2+4((g u2)/2+g Cos[\[Theta]])^2)))
+ Sin[\[Theta]]Log[(1+xm1 xp1-(xm1+xp1) Cos[\[Theta]]
-I (xm1-xp1) Sin[\[Theta]])/(1+xm1 xp1-(xm1+xp1) 
Cos[\[Theta]]+I (xm1-xp1) Sin[\[Theta]])]
(PolyGamma[1+M/2+(I g u2)/2-I g Cos[\[Theta]]]
+PolyGamma[1+M/2-(I g u2)/2+I g Cos[\[Theta]]])
+Sin[\[Theta]]Log[(1+xm1 xp1+(xm1+xp1) Cos[\[Theta]]
-I (xm1-xp1) Sin[\[Theta]])/(1+xm1 xp1+(xm1+xp1) 
Cos[\[Theta]]+I (xm1-xp1) Sin[\[Theta]])]
(PolyGamma[1+M/2+ (I g u2)/2+I g Cos[\[Theta]]]
+PolyGamma[1+M/2-(I g u2)/2-I g Cos[\[Theta]]]),{\[Theta],0, Pi/2}]);
\end{verbatim}
}
Here the definitions  are the same as above but $Q=1$, and $x_1^\pm$ are in the string region, that is $x_1^\pm=x_s(u_1\pm{i\ov g})$, while $x_2^\pm$ are in the mirror region, that is $x_2^\pm=x(u_2\pm{i\ov g}M)$. Then, 
\eqref{psi21b} is written as
{\smaller
\begin{verbatim}
\[CapitalSigma]\[CapitalPsi]21[u1_, Q_, u2_, M_, g_] := 
\[CapitalPsi]21[xs[u1 + I*(Q/g)], xs[u1 - I*(Q/g)], u2, g, M]; 
\end{verbatim}
}
The last term in the formula \eqref{lastsm} for the  improved string-mirror dressing phase is given by
{\smaller
\begin{verbatim}
\[CapitalSigma]lastsm[u1_, Q_, u2_, M_, g_] = (1/(2*I))*Log[(xs[u1 - I*(Q/g)] 
- x[u2 + I*(M/g)])*(xs[u1 - I*(Q/g)] - 1/x[u2 - I*(M/g)])*((xs[u1 + I*(Q/g)] 
- 1/x[u2 - I*(M/g)])/((xs[u1 + I*(Q/g)] - x[u2 + I*(M/g)])*(xs[u1 - I*(Q/g)] 
- 1/x[u2 + I*(M/g)])^2))]; 
\end{verbatim}
}
The total improved dressing phase \eqref{drsm} in the string-mirror region is thus given by the sum of the three terms
{\smaller
\begin{verbatim}
\[CapitalSigma]Ism[u1_, 1, u2_, M_, g_] := \[CapitalTheta]sm[u1, 1, u2, M, g] 
+ \[CapitalSigma]\[CapitalPsi]21[u1,1,u2,M,g] + \[CapitalSigma]lastsm[u1,1,u2,M,g];
\end{verbatim}
}

The  total contribution \eqref{psi21mm} of $\Psi(x_2,x_1)$-functions in the mirror-mirror region  is
{\smaller
\begin{verbatim}
\[CapitalPsi]21m[xp1_, xm1_, u2_, g_, M_] := 
(g/(2*Pi))*NIntegrate[Sin[\[Theta]]*(2*\[Theta] - 2*Pi 
- I*Log[(1 - 1/(E^(I*\[Theta])*xm1))/(1 - E^(I*\[Theta])/xm1)] - 
I*Log[(1-xp1/E^(I*\[Theta]))/(1-E^(I*\[Theta])*xp1)])*((-2*M)/(M^2+4*((g*u2)/2
 - g*Cos[\[Theta]])^2) +  PolyGamma[1 + M/2 + (I*g*u2)/2 - I*g*Cos[\[Theta]]] 
 + PolyGamma[1 + M/2 - (I*g*u2)/2 + I*g*Cos[\[Theta]]]) + 
Sin[\[Theta]]*(-2*\[Theta]-I*Log[(1+E^(I*\[Theta])/xm1)/(1+1/(E^(I*\[Theta])*xm1))]- 
 I*Log[(1+E^(I*\[Theta])*xp1)/(1+xp1/E^(I*\[Theta]))])*((-2*M)/(M^2+4*((g*u2)/2
 +g*Cos[\[Theta]])^2) + PolyGamma[1 + M/2 + (I*g*u2)/2 + I*g*Cos[\[Theta]]] + 
 PolyGamma[1 + M/2 - (I*g*u2)/2 - I*g*Cos[\[Theta]]]),  {\[Theta], 0, Pi/2}] + 
 (I/2)*(LogGamma[1+(I*(g*u2+I*M))/2-I*g] -LogGamma[1-(I*(g*u2+I*M))/2+I*g]+ 
 LogGamma[1+(I*(g*u2-I*M))/2-I*g] -LogGamma[1-(I*(g*u2-I*M))/2+I*g]); 
\end{verbatim}
}
The  total contribution \eqref{psi12mm} of $\Psi(x_1,x_2)$-functions  is given by
{\smaller
\begin{verbatim}
\[CapitalPsi]12m[u1_, xp2_, xm2_, g_, M_] := 
(g/(2*Pi))*NIntegrate[Sin[\[Theta]]*(-2*\[Theta] + 2*Pi 
+ I*Log[(1 - 1/(E^(I*\[Theta])*xm2))/(1 - E^(I*\[Theta])/xm2)] + 
I*Log[(1-xp2/E^(I*\[Theta]))/(1-E^(I*\[Theta])*xp2)])*((-2*M)/(M^2+4*((g*u1)/2 
- g*Cos[\[Theta]])^2) + PolyGamma[1 + M/2 + (I*g*u1)/2 - I*g*Cos[\[Theta]]] 
+ PolyGamma[1 + M/2 - (I*g*u1)/2 + I*g*Cos[\[Theta]]]) + Sin[\[Theta]]*(2*\[Theta] 
 + I*Log[(1 + E^(I*\[Theta])/xm2)/(1 + 1/(E^(I*\[Theta])*xm2))] 
 + I*Log[(1+E^(I*\[Theta])*xp2)/(1+xp2/E^(I*\[Theta]))])*((-2*M)/(M^2+4*((g*u1)/2 
 + g*Cos[\[Theta]])^2) +PolyGamma[1 + M/2 + (I*g*u1)/2 + I*g*Cos[\[Theta]]] 
 + PolyGamma[1 + M/2 - (I*g*u1)/2 - I*g*Cos[\[Theta]]]), {\[Theta], 0, Pi/2}] - 
(I/2)*(LogGamma[1+(I*(g*u1+I*M))/2-I*g]-LogGamma[1-(I*(g*u1+I*M))/2+I*g] + 
 LogGamma[1 + (I*(g*u1 - I*M))/2 - I*g]-LogGamma[1 - (I*(g*u1 - I*M))/2 + I*g]); 
\end{verbatim}
}

$\Psi_{12}^{\rm mm}$ and $\Psi_{21}^{\rm mm}$ functions in \eqref{psimms} are
{\smaller
\begin{verbatim}
\[CapitalSigma]\[CapitalPsi]12m[u1_, Q_, u2_, M_, g_] 
:= \[CapitalPsi]12m[u1, x[u2 + I*(M/g)], x[u2 - I*(M/g)], g, Q]; 
\[CapitalSigma]\[CapitalPsi]21m[u1_, Q_, u2_, M_, g_] 
:= \[CapitalPsi]21m[x[u1 + I*(Q/g)], x[u1 - I*(Q/g)], u2, g, M]; 
\end{verbatim}
}
Finally, the last term \eqref{lastmm} is given by 
{\smaller
\begin{verbatim}
\[CapitalSigma]lastmm[u1_, Q_, u2_, M_, g_] := 
(1/I)Log[(I^Q/I^M)(Gamma[M-(I/2)(g*u1-g*u2+I(Q-M))]/Gamma[Q+(I/2)(g*u1-g*u2+I(Q-M))])*
 ((1-1/(x[u1+I*(Q/g)]*x[u2-I*(M/g)]))/(1-1/(x[u1-I*(Q/g)]*x[u2+I*(M/g)])))*
      ((x[u1+I*(Q/g)]/x[u1-I*(Q/g)])*(x[u2-I*(M/g)]/x[u2+I*(M/g)]))^(1/2)]; 
\end{verbatim}
}
Thus, the improved dressing phase \eqref{phmm} in the mirror-mirror region is 
{\smaller
\begin{verbatim}
\[CapitalSigma]Imm[u1_, Q_, u2_, M_, g_] :=
 \[CapitalTheta]mm[u1,Q,u2,M,g] +\[CapitalSigma]\[CapitalPsi]12m[u1,Q,u2,M,g] 
 +\[CapitalSigma]\[CapitalPsi]21m[u1,Q,u2,M,g]+\[CapitalSigma]lastmm[u1,Q,u2,M,g]; 
\end{verbatim}
}


\end{document}